\documentclass[twocolumn]{aastex701}

\newcommand{\pc}[2]{\parbox[t]{#1}{#2\strut}}

\usepackage{amsmath,amssymb,amsfonts,amsthm}
\usepackage{graphicx}
\usepackage{bm}
\makeatletter
\newcommand\algorithmname{Algorithm}
\newcounter{algorithm}
\def\fps@algorithm{tbp}
\def\ftype@algorithm{4}
\def\ext@algorithm{loa}
\def\fnum@algorithm{\algorithmname~\thealgorithm}
\newenvironment{algorithm}{\@float{algorithm}}{\end@float}
\newenvironment{algorithm*}{\@dblfloat{algorithm}}{\end@dblfloat}
\makeatother
\usepackage{algorithmic}
\usepackage{booktabs}
\usepackage{longtable}
\usepackage{xcolor}

\newcommand{\vect}[1]{\bm{#1}}
\newcommand{\mat}[1]{\mathbf{#1}}

\newcommand{\dd}{\mathrm{d}}
\newcommand{\reals}{\mathbb{R}}

\newcommand{\passmark}{$\checkmark$}
\newcommand{\partialmark}{$\sim$}
\newcommand{\failmark}{$\times$}

\shorttitle{NestyNet. IV. Laws Chosen by Nothing in Advance}
\shortauthors{Ibata et al.}

\begin{document}

\title{NestyNet. IV. Laws Chosen by Nothing in Advance}

\author[0000-0002-3292-9709]{Rodrigo Ibata}
\affiliation{Universit\'e de Strasbourg, CNRS, Observatoire astronomique de Strasbourg, UMR 7550, F-67000 Strasbourg, France}
\email[show]{rodrigo.ibata@astro.unistra.fr}

\author[0000-0001-8392-3836]{Wassim Tenachi}
\affiliation{Mila - Quebec Artificial Intelligence Institute}
\affiliation{D\'epartement de physique, Universit\'e de Montr\'eal}
\affiliation{Ciela - Montreal Institute for Astrophysical Data Analysis and Machine Learning}
\email{wassim.tenachi@umontreal.ca}

\author[0000-0002-8788-8174]{Foivos Diakogiannis}
\affiliation{Technology, Commonwealth Scientific and Industrial Research Organisation (CSIRO), Kensington, WA 6151, Australia}
\email{Foivos.Diakogiannis@data61.csiro.au}

\author[0000-0003-1559-1053]{Neil Ibata}
\affiliation{Department of Human Evolutionary Biology, Harvard University, Cambridge, MA, USA}
\email{neilibata@fas.harvard.edu}

\author[0009-0008-7455-1880]{Anirudh Shankar}
\affiliation{Universit\'e de Strasbourg, CNRS, Observatoire astronomique de Strasbourg, UMR 7550, F-67000 Strasbourg, France}
\email{anirudh.shankar@astro.unistra.fr}

\begin{abstract}
Differential-equation (DE) discovery tends to break down precisely where
much of physics begins.  Fields are coupled, governing laws are nonlinear in
the state, amplitudes, coordinates, or operators of interest, yet
derivatives must remain consistent across fields, channels, and
differentiation orders. NestyNet-DE addresses this via two complementary DE
search strategies, both employing analytic derivatives from segmented neural
surrogates: a sparse-library route linear in the outer coefficients, and a
new operator-factorized route that searches directly over equation structure
rather than a fixed library, recovering compositional laws the sparse route
misses. The recovered laws can be strongly nonlinear in states, fields,
coordinates, and their couplings.  The framework also handles multi-dataset
shared-support discovery, complex and vector laws, and Hamiltonian discovery
from phase-space trajectories. Beyond a law's form, the same data yield its
geometry, the Lie point symmetries of the recovered equation.  We apply the
pipeline to 30 years of daily ephemerides of $308$ main-belt asteroids and
recover the reduced-Kepler hierarchy (areal law, inverse-square force, and
reduced Hamiltonian), where a discovered rotational symmetry fixes the
centrifugal coefficient rather than fitting it.  We also present a benchmark
of 57 real-valued ODEs and 26 complex-valued systems, including the
Schr\"odinger and Dirac equations, with Maxwell's equations as a coupled
vector-PDE case study.  Here, discovery is not shackled to a fixed library,
nor does it end at the equation. It composes laws that no dictionary
anticipated, and closes the loop from data, to a law chosen by nothing in
advance, to the geometry that explains and integrates it.
\end{abstract}


\section{Introduction}
\label{sec:introduction}

Discovering governing differential equations directly from data is one of
the most ambitious forms of machine-assisted scientific inference.  The goal
is to recover a compact law
that a physicist can inspect, criticize, refute, reuse for prediction, or
even develop further into an actual scientific theory.  Achieving this for
simple scalar systems is already useful, but it becomes much more important,
and much more difficult, for coupled fields, complex amplitudes, and
structured Hamiltonian dynamics.  Given trajectories or fields $\vect{u}(x)$
sampled over one or more independent variables $x$ (time, space, or both),
we seek to infer an operator relation
\begin{equation}
\mathcal{L}[\vect{u}](x) = 0
\end{equation}
that is predictive on held-out data and intelligible to a physicist.  Sparse
library methods such as the Sparse Identification of Nonlinear Dynamics
(SINDy)~\citep{Brunton2016} and PDE-FIND~\citep{Rudy2017} made this problem
tractable by restricting attention to equations that are linear in unknown
coefficients.  A direct consequence is that every library term must be fixed
in advance, and none may carry a free nonlinear parameter of its own, so a
term such as $A\sin(\omega t)$ can contribute its amplitude $A$ as a linear
coefficient, but its frequency $\omega$ must be committed to beforehand.
Another part of the difficulty is in the interface
between data and library.  Derivatives must be estimated accurately, coupled
laws must be handled coherently, and physically meaningful shared structure
should be accessible to analysis and not hidden inside black-box fits.

In practice, one of the dominant limitations to differential equation
discovery is the quality of the derivatives supplied to it.  Finite
differences and interpolation can both be effective in favorable regimes,
but they inevitably inject approximation error into the library.  
Weak-form methods such as
WSINDy~\citep{Messenger2021WSINDy,Messenger2021WSINDyPDE} instead integrate
the candidate terms against compactly supported test functions, avoiding
pointwise differentiation of noisy data at the cost of tying recovery to the
choice of test functions and supports, while leaving the fixed-library
constraint in place.  The weak form also removes precisely the pointwise
derivative features that the operator-factorized search and the symmetry
machinery introduced below consume.  The difficulty becomes especially acute
for higher derivatives, mixed partials, sparse or noisy measurements, and
multi-field systems in which several derivative channels must remain
mutually consistent.  As we showed in \citet[hereafter
Paper~I]{NestyNet2026a}, standard multilayer perceptron (MLP) surrogates
trained with first-order methods can fit function values well while yielding
substantially less accurate derivatives.  There, NestyNet surrogates
improved accuracy over Adam-trained~\citep{KingmaBa2015} MLPs by median
factors of order $10^{3}$ for function values and $10^{2}$--$10^{3}$ for
first and second derivatives and operator probes.  For equation discovery,
that gap can be the difference between recovering the right support and
fitting a numerically plausible but physically wrong law.  In
\citet[hereafter Paper~III]{NestyNet2026b} we demonstrated that the accurate
NestyNet derivatives enable exact equation recovery at a level not
previously reported in symbolic regression (SR). 
%

Our aim is to develop an algorithm capable of discovering many of the
flagship differential equations of physics from data, while also being
flexible enough to propose new, unexpected symbolic expressions. The
machinery should allow for coupled vector equations, such as Maxwell's, and
for complex-valued differential equations, such as Schr\"odinger's.

Several lines of work pursue related goals.  The sparse-library idea
has been extended to parametric/shared-sparsity settings~\citep{Rudy2019}
and, through SINDy-PI~\citep{Kaheman2020}, to implicit and rational
dynamics, though every candidate term must still be committed in advance.
Across these variants, LASSO~\citep{Tibshirani1996} and sequential
thresholded least squares are the typical sparsifying solvers.  Autoencoders
have been employed to jointly discover coordinates and governing
equations~\citep{Champion2019}.
Open-form searches relax the fixed library by evolving or
learning tree-structured PDE terms~\citep{Chen2022SGAPDE,Du2024DISCOVER} or
by searching ODE space with formal grammars~\citep{Yu2025GODE}.
Black-box dynamics learners such as Neural
ODEs~\citep{Chen2018} parameterize the right-hand side of
$\dot z = f_\theta(z,t)$ directly, while Hamiltonian and Lagrangian neural
networks~\citep{Greydanus2019,Cranmer2020HNN} enforce energy or variational
structure by construction, but in both cases the learned object remains a
neural network rather than an explicit law.  A complementary deep SR
approach instead recovers explicit analytical expressions directly
from data, for example via units-constrained reinforcement
learning~\citep{Tenachi2023PhySO}, via transformers pretrained on synthetic
systems~\citep{dAscoli2024ODEFormer}, and, for datasets that share a common
functional form, class symbolic regression~\citep{Tenachi2024ClassSR}.

A distinct and more recent thread brings \emph{symmetry} to bear on this
problem.  On the discovery side, symmetry-informed sparse regression uses a \emph{known}
invariance to prune the candidate library~\citep{Yang2024SymmetryInformed},
and a related approach detects a predefined symmetry from data to reduce the
library for PDE identification~\citep{EqOD2026}.  On the geometry side,
symmetry-discovery methods learn Lie-group generators from
data~\citep{Yang2023LieGAN} and conservation-law discovery extracts
invariants from trajectories~\citep{Liu2021AIPoincare}.  These characterize
a dataset or flow rather than a discovered symbolic equation, while
classical Lie analysis~\citep{Olver1993} begins from the opposite end, 
from a known equation.

Our contribution occupies the complementary position.  We discover the Lie
point (and, in phase space, symplectic) symmetries of the equation we
have just proposed.  That geometry then works in both directions.  As an
input it arbitrates among candidates and manufactures library atoms, and as
an output it integrates the recovered law and, for the reduced Kepler showcase
(\S\ref{sec:kepler-showcase}), derives a structural coefficient rather than 
just fitting it.


\section{Design}
\label{sec:design}

NestyNet-DE has a two-tiered design.  The first tier implements differential
equation discovery by extending the SR engine of Paper~III 
with a Sequential Thresholded Least Squares
(STLSQ)~\citep{Brunton2016} module.  The NestyNet surrogate
can be fit to trajectory or field data, allowing the full
AST-encoded library and its derivatives to be evaluated in closed form.
STLSQ then selects a sparse set of active terms, and the surviving law can
be refined with VarPro~\citep{Golub1973,Golub2003}.  The first tier thus
inherits the structure of SINDy and PDE-FIND, but replaces their numerically
differentiated candidate libraries with terms evaluated analytically from a
frozen and accurately trained surrogate.

Although the backbone of this first tier is linear in the outer
coefficients, the candidate terms themselves can encode strongly nonlinear
state dependences, singular coordinate factors, coupled complex amplitudes,
and vector-calculus operators.  The same AST representation developed in
Paper~III supports shared-support multi-dataset discovery, complex-valued
laws handled through coupled-real components, vector-valued laws built from
a vector-calculus term grammar with coefficient tying and cross-equation
shared constants, and Hamiltonian discovery from phase-space data.  The
candidate library of this tier comes in two flavors: a standard dictionary
for the scalar cases, extended with inverse-coordinate atoms only when a
case declares a singular-origin domain in its structural metadata, and a
metadata-driven \emph{class library} that specializes the candidate set to a
problem's structural class (used for the structured complex cases,
\S\ref{sec:complex-laws}).  Both are informed only at the level of structure
(the declared class of the problem) and never by the terms of the actual
target equation.  This distinguishes them from oracle libraries that
presuppose the answer.

The second tier, a new operator-factorized route that searches over typed
factorized balances, is adapted from the factorized symbolic search of 
Paper~III and detailed in \S\ref{sec:operator-factorized}.  It fixes a grammar
rather than a menu.  
\section{AST Representation for Equation Libraries}
\label{sec:ast}

NestyNet-DE represents candidate library terms and discovered equations
using the same native AST substrate introduced in Paper~III: binary
operators (Add, Mul, Pow), unary operators (Log, Exp, and the trigonometric
family), input coordinates, fixed constants, and atom nodes whose input
expressions may themselves be arbitrary ASTs.  For equation discovery, the
critical additions are the surrogate field leaves $U_m$,
$\partial_{x_a} U_m$, and $\partial_{x_a x_b} U_m$, which evaluate the
$m$-th component of the trained surrogate and its first and second partial
derivatives along axes $x_a$ and $x_b$.  Optional named field atoms map
semantic objects such as electric and magnetic fields $\vect{E}$ and
$\vect{B}$ to output components, allowing vector-calculus libraries while
retaining a scalar backend.  All library terms inherit the Paper~III
compilation layer, with analytic chain-rule input gradients and Hessians flowing
through the AST, and Jacobian--vector and vector--Jacobian products feeding the
Levenberg--Marquardt~\citep{Levenberg1944,Marquardt1963} solver.  
%

\section{Differential Equation Discovery}
\label{sec:ode}

\subsection{Linear-in-Coefficients Formulation}

We consider a real-valued state vector $\vect{u}(x)\in\reals^{M}$, where
complex fields are represented by stacking real and imaginary parts and
vector fields are represented by stacking components, and where
$x\in\reals^{n}$ collects one or more independent variables (e.g.,
$t,x,y,z$).  We search for equations that are linear in unknown coefficients
$c_{mk}$,
\begin{equation}
\mathcal{L}_m[\vect{u}] + \sum_{k=1}^{K} c_{mk}\,\phi_k(x,\vect{u},\partial\vect{u},\partial^2\vect{u}) = 0,
\qquad m=1,\ldots,M,
\label{eq:ode_library}
\end{equation}
where $\phi_k$ are candidate library terms.  The ``anchor'' operator
$\mathcal{L}_m$ is chosen from a small family (typically $\partial_{x_a}u_m$
for first-order dynamics or $\partial_{x_a x_a}u_m$ for wave-like
equations), so that the remaining library terms explain the residual.

Given a frozen neural surrogate $\hat{\vect{u}}(x)$, both $\hat{\vect{u}}$
and its derivatives are known analytically. We construct a feature matrix
$\mat{\Phi}$ with entries $\Phi_{ik}=\phi_k(x_i,\hat{\vect{u}}(x_i),\ldots)$
and per-equation targets $b^{(m)}_i=-\mathcal{L}_m[\hat{\vect{u}}](x_i)$,
then solve $\mat{\Phi}\,\vect{c}^{(m)}=\vect{b}^{(m)}$ for each
equation. All library terms are represented as ASTs built from neural atoms,
ensuring consistent evaluation and derivatives throughout.

The standard scalar library includes a constant term~$1$, powers of
state components ($u_j$, $u_j^2$, $u_j^3$, \ldots), cross-terms
($u_i u_j$, $x_a u_j$, $u_j \,\partial_{x_a}u_k$), first derivatives
$\partial_{x_a}u_j$ along any axis $x_a$, and second derivatives or
cross-partials $\partial_{x_a x_b}u_j$.  For cases whose structural
metadata declares a singular-origin domain (a coordinate singularity at
the origin, as in radial problems), the library additionally admits the
inverse-coordinate atoms $u_j/x_a$, $u_j/x_a^{2}$, and
$(\partial_{x_a}u_j)/x_a$ (this gate reads only the declared flag, never
the target equation).  The same AST machinery also accommodates additional term families,
such as the power-law and exponential templates later refined by
VarPro (\S\ref{sec:varpro}), and any such terms share the uniform
evaluation and differentiation path of the standard entries.

\subsection{STLSQ: Sequential Thresholded Least Squares}

Following SINDy and PDE-FIND~\citep{Brunton2016,Rudy2017,Rudy2019}, we use
STLSQ for sparse coefficient selection, which alternates between ridge
regression and hard thresholding of coefficients with magnitude below
$\lambda$, iterating until the support converges.  The threshold $\lambda$
controls sparsity, with larger values yielding simpler equations.  We use
STLSQ rather than LASSO because iterative hard thresholding is both more
interpretable and more robust when the feature matrix is moderately
ill-conditioned, as is typical for derivative-based libraries.

Although STLSQ is linear in coefficients, the recovered law need not be.  It
can still contain trigonometric, polynomial, rational, singular-coordinate,
amplitude-dependent, complex-cubic, or vector-operator structure, but the
regression remains linear in the outer coefficients multiplying those
candidate terms.  

\subsection{Operator-Factorized Search (OFS)}
\label{sec:operator-factorized}

When the correct law is a composition that no fixed library column
represents, the second tier searches directly over equation
\emph{structure}.  The law is treated as a \emph{balance} in implicit form,
so that its terms sum to zero, and instead of selecting columns from a fixed
library (as STLSQ does) the route searches over \emph{factorized} balances
\begin{equation}
\mathcal{L}_m[\vect{u}] + \sum_j g_j(z_j)\,\phi_j = 0 ,
\label{eq:factorized_balance}
\end{equation}
in which each term is a product of a few \emph{typed} factors, namely a
fixed \emph{carrier} $\phi$ (a low-order monomial such as $u$ or
$\partial_x u$, or a vector operator $\nabla\!\cdot$, $\nabla\!\times$,
$\nabla^2$) multiplied by a discovered coefficient $g(z)$, a univariate
function of a single argument $z$.  Choosing $z$ to be a coordinate yields a
coordinate-only coefficient $a(x)$, while choosing it to be a state yields a
state-only nonlinearity $b(u)$.  Given a typed skeleton, the linear
coefficients are fit by robust least squares, optionally on top of the terms
STLSQ has already selected, and the surviving candidate is validated by
forward rollout.  The conceptual shift is from sparse selection among
pre-existing library columns to sparse selection of typed operator
structure.  This expands the ``factorized symbolic search'' method explained
in Paper~III, running either as the first line (with STLSQ disabled) or as a
rescue lane when the STLSQ fit leaves a high residual or is ill-conditioned.
Algorithm~\ref{alg:operator-factorized-de} gives the two-phase procedure, in
which cheap coefficient-free typed skeletons are tried first and an
archive-guided search then mutates the survivors.

\begin{algorithm}[t]
\caption{Operator-factorized DE discovery: the DE-discovery instantiation of
factorized symbolic search. The residual-basin archive, fingerprinting, and
upper-confidence-bound (UCB) selection are inherited from Paper~III. The steps
below highlight what is specific to DE discovery.}
\label{alg:operator-factorized-de}
\begin{algorithmic}[1]
\REQUIRE Surrogate(s) $u(\mathbf{x})$ with analytic derivatives on fit/probe splits, library $\Phi$, mapping battery $\mathcal{M}$ (the coefficient-function families for $g$, e.g.\ affine, power, exponential, logarithm, and rational), operator set $\mathcal{O}=\{\mathrm{grad},\mathrm{div},\mathrm{curl},\nabla^{2}\}$, optional units.
\STATE Read the leading-order \emph{anchor} $A$ (e.g.\ $\partial_{t}u$) and features $(\mathbf{x},u,\partial u,\dots)$ from the surrogate.  The target is the implicit balance $A+\sum_{k}c_{k}\phi_{k}=0$.
\STATE \textbf{Tier~1 (sparse library):} group-sparse STLSQ over $\Phi$ with support tied across datasets, optionally VarPro-refined, then \RETURN if its residual and conditioning gates pass.
\STATE \textbf{Tier~2 (operator-factorized search, first line when STLSQ is off, else a rescue lane):}
\STATE Initialize a residual-fingerprint archive whose \emph{basins} group candidates that share a residual fingerprint, each basin keeping its best (elite) members.
\STATE \textbf{Phase~1 (shallow typed balances):}
\FOR{each shallow coefficient-free skeleton over $(\mathbf{x},u,\partial u,\dots)$ and $\mathcal{O}$, unit-filtered}
  \STATE Fit the best mapping in $\mathcal{M}$ and dataset-tied coefficients, score against $A$ across trajectories, update basin elites, and stop on a dimensionally consistent fit.
\ENDFOR
\STATE \textbf{Phase~2 (archive-guided search):}
\FOR{$t=1,\ldots,n_{\mathrm{iter}}$}
  \STATE Pick a basin by UCB, trading off log-MSE gain against novelty, then form an action slate (build, residual, closure, sparse-combo, inverse-/hole-repair) under dimensional, depth, residual, and domain-fragility gates.
  \STATE For closure/sparse-combo actions, refit and backward-prune the coefficient head on $A$ across datasets to few terms, and optionally refine skeletons.
  \STATE Update route statistics and soft-restart on stalls.
\ENDFOR
\STATE \textbf{Validation:} roll out (integrate) the top candidates and rank them by integration success, domain safety, integration/probe error, then size.
\RETURN Top-$k$ balances, each with skeleton, dataset-tied coefficients, mapping metadata, dimensional signature, and integration diagnostics.
\end{algorithmic}
\end{algorithm}

\subsection{Coupled Systems and Multi-Task Discovery}
\label{sec:multitask}

Many scientific laws appear as coupled systems, with multiple state
components, multiple equations, and evidence spread across multiple datasets
(different initial conditions or experiments).  We treat each
equation--dataset pair as a ``task'' and require that all tasks share the
same set of active library terms while allowing each task its own
coefficient values.  This brings the principle of class symbolic
regression~\citep{Tenachi2024ClassSR} to differential-equation discovery,
where the functional form of the law is universal (e.g.\ the same terms
appear in every equation or dataset) even when the numerical coefficients
differ.  Operationally, we collect the per-task coefficients into a matrix
$\mat{C}\in\reals^{T\times K}$ and threshold term~$k$ when the column norm
across all tasks is small, $\|\mat{C}_{:,k}\|_2<\lambda$, then re-solve on
the surviving support.  In vector problems, a single task may itself be a
component-stacked vector regression (\S\ref{sec:vector_tying}). The same
group thresholding can then be applied across datasets and/or across
multiple coupled vector laws to recover a universal operator support (e.g.,
common curl/div structure) while allowing coefficients to vary when desired.
This generalizes the parametric shared-sparsity idea~\citep{Rudy2019} from
parameter sweeps to a broader task space that includes different equations,
datasets, vector laws, and coupled-field components.

\subsection{Complex-Valued Laws via Real Decomposition}
\label{sec:complex-laws}

Complex-valued equations are handled by representing each field
$\psi=u+iv$ as paired real surrogate outputs.  A single complex field therefore
appears as a two-component real system, and coupled complex fields become
higher-dimensional real systems.  This allows the same analytic-derivative and
sparse-regression machinery to operate unchanged while still exposing the
physical structure of the original complex law.

On top of this real decomposition we support structured complex library
terms such as $|\psi|^2\psi$, complex products, and Laplacians of $\psi$,
expanded into coupled real and imaginary AST components.  For the complex
benchmark these terms are assembled into a \emph{metadata-driven class
library} rather than a single generic dictionary.  The candidate set for each
problem is built from its class metadata (the number of complex components,
the number of spatial axes, the anchor differential order, and the declared
complex-operation set) and populated
with the structure characteristic of that class (fields, modulus
nonlinearities $|\psi|^2\psi$, pairwise field products, coordinate--field
products, first and second derivatives and Laplacians, and trigonometric
carriers).  STLSQ is then run over a small set of such class-level library
variants and the best is selected by its training residual with a parsimony 
penalty.

This candidate set is fixed by the structural metadata alone, never by the
target equation's own terms, so it is structurally distinct from an oracle
library that would presuppose the answer.  We therefore refer to the complex
first line as \emph{class-library STLSQ} to distinguish it from the
standard-library STLSQ used on the scalar cases.  When such a
physics-motivated complex library is used, we can additionally validate
coefficient patterns associated with Hermitian and anti-Hermitian operators
in equations of the form $i\partial_t\psi=\mathcal{L}[\psi]$.  This is the
route by which the framework targets Schr\"odinger, Gross--Pitaevskii,
Ginzburg--Landau, Dirac, and coupled complex-field models without a separate
complex-valued symbolic backend.

\subsection{Vector-Calculus Grammar and Coefficient Tying}
\label{sec:vector_tying}

Standard SINDy/PDE-FIND libraries are typically written as scalar
partials and products.  To move beyond scalar PDEs, we introduce a 
vector term layer that generates structured candidates such as
$\nabla\!\cdot$, $\nabla\!\times$, $\nabla$, and $\nabla^2$ from named vector
fields, with higher-level operations (dot, cross, advection
$(\vect{v}\cdot\nabla)\vect{u}$) built compositionally from these primitives.
Each vector expression expands into its component scalar ASTs, so the
underlying evaluator remains purely scalar while preserving the vector
structure for coefficient tying.

Given a single vector equation of dimension $d$, we enforce a single
coefficient vector shared across
components by stacking component residuals into one regression problem.
Let $\vect{\phi}_k(x)\in\reals^{d}$ denote vector-valued library terms and
$\vect{\mathcal{L}}[\vect{u}]\in\reals^{d}$ the vector anchor.  We solve
\begin{equation}
\begin{bmatrix}
\mat{\Phi}_x\\ \mat{\Phi}_y\\ \vdots
\end{bmatrix}\vect{c}
=
-
\begin{bmatrix}
\vect{b}_x\\ \vect{b}_y\\ \vdots
\end{bmatrix},
\end{equation}
where each block corresponds to a component (e.g., $x,y,z$) and shares the
same coefficients $\vect{c}$. This stacking enforces the physical
requirement that a vector law have consistent constants across all spatial
components, a single wave speed, a single viscosity, and so on.

\subsection{Cross-Equation Coefficient Sharing}
\label{sec:coeffsharing}

We further support simultaneous discovery of multiple vector equations
(e.g., coupled evolution laws for $\vect{E}$ and $\vect{B}$), with shared
physical constants encoded by coefficient-sharing groups.  Writing each
constrained coefficient as $c^{(t)}_{k}=s^{(t)}_{k}\,g_{j}$, for a global
coefficient $g_j$ and a fixed scale $s^{(t)}_{k}$ that absorbs signs and
unit conventions, collapses the per-task regressions
$\mat{\Phi}^{(t)}\vect{c}^{(t)}=\vect{b}^{(t)}$ into a single reduced
regression $\mat{\Phi}_{\rm red}\vect{g}=\vect{b}_{\rm all}$, on which STLSQ
runs directly.  Because the number of global coefficients can be far smaller
than the number of task coefficients, the solve favors explanations that
reuse the same constant across many tasks, for example a single wave speed
shared between coupled vector laws and across experimental realizations.
Assigning each coefficient its own group recovers the unconstrained
formulation.

\subsection{Hamiltonian Discovery}
\label{sec:hamiltonian}

NestyNet-DE also discovers Hamiltonian dynamics from phase-space
trajectories $(q(t),p(t))$, automatically preserving symplectic
structure. Parameterizing $H(q,p)$ as a polynomial, optionally with the
mechanical split $H=T(p)+V(q)$, makes Hamilton's equations
$\dot q=\partial H/\partial p$ and $\dot p=-\partial H/\partial q$ linear in
the coefficients, so the same STLSQ machinery recovers $H$ from the stacked
$(\dot q,\dot p)$ regression, with the multi-trajectory group thresholding
of \S\ref{sec:multitask} carrying over unchanged.

\subsection{Dimensional Analysis in the DE Setting}
\label{sec:units:de}

Dimensional analysis is central to NestyNet-DE.  The underlying
machinery is inherited unchanged from Paper~III, where we introduced a
two-level units engine, with local dimensional gates for AST operations together
with a global ``Buckingham--Sudoku'' constraint propagation system.  
What matters for the present contribution is how that machinery is instantiated 
for differential-equation libraries, derivative atoms, vector operators, and 
shared coefficients.

For DE discovery the basic unit-carrying objects are the surrogate field
components and their analytic derivatives.  If the $m$th field component has
dimension $[u_m]$ and axis $x_a$ has dimension $[x_a]$, then the canonical DE
atoms carry
\begin{equation}
\begin{aligned}
{[U_m]} &= [u_m],\\
{[\partial_{x_a}U_m]} &= [u_m]-[x_a],\\
{[\partial_{x_a x_b} U_m]} &= [u_m]-[x_a]-[x_b] \, .
\end{aligned}
\label{eq:de_units_atoms}
\end{equation}
This is the DE-specific bridge between the generic AST units engine and the
analytic-derivative feature tables used in Eq.~\eqref{eq:ode_library}.

The anchor operator in Eq.~\eqref{eq:ode_library} then fixes the admissible
dimension of every candidate contribution in equation $m$,
\begin{equation}
[c_{mk}] + [\phi_k] = [\mathcal{L}_m[\vect{u}]].
\label{eq:de_units_balance}
\end{equation}
Terms that violate this balance are rejected before regression.  Thus
dimensional analysis is not an after-the-fact sanity check but one
of the main filters that keeps the DE library physically coherent before any
sparse solve is attempted.

Complex and vector terms inherit these units component-wise.  With $\psi=u+iv$
giving $[u]=[v]=[\psi]$, a term such as $|\psi|^2\psi$ carries $3[\psi]$ and
$\partial_x^2\psi$ carries $[\psi]-2[x]$, while $\nabla\phi$,
$\nabla\!\cdot\vect{F}$, $\nabla\!\times\vect{F}$, and $\nabla^2\phi$ take the
expected derivative-shifted dimensions.  The same Buckingham--Sudoku machinery
therefore prunes coupled-real and vector systems that are algebraically legal but
physically inconsistent, and the vector-calculus grammar of \S\ref{sec:vector_tying} is
filtered by unit compatibility as well as syntax.

This becomes especially important once coefficients are tied across components
or across multiple equations.  A shared coefficient group is admissible only if
all tied occurrences induce the same required coefficient dimension under
Eq.~\eqref{eq:de_units_balance}.  Otherwise the tie is physically meaningless
and is rejected.

\section{Variable Projection Refinement}
\label{sec:varpro}

VarPro~\citep{Golub1973,Golub2003} separates linear from nonlinear
parameters, and we exploit this separation to push the current pipeline
beyond a bare sparse linear solve without sacrificing interpretability.
Once a promising support has been identified, selected terms can carry
nonlinear shape parameters and those parameters can be optimized while the
linear coefficients are analytically eliminated.  In practice this is what
lets the structured first-tier route reach fractional powers and
exponentials without turning the entire discovery problem into an
unconstrained fit.

\subsection{Two-Phase Optimization}

Given fixed nonlinear parameters $\theta_{\rm nl}$ (e.g., template
exponents), the linear coefficients $\vect{c}$ are determined analytically
by
\begin{equation}
\hat{\vect{c}}(\theta_{\rm nl}) = (\mat{\Phi}(\theta_{\rm nl})^\top \mat{\Phi}(\theta_{\rm nl}) + \mu \mat{I})^{-1} \mat{\Phi}(\theta_{\rm nl})^\top \vect{y} .
\end{equation}
Here $\vect{y}$ stacks the regression targets and $\mu>0$ is a small ridge
parameter. This eliminates the linear unknowns entirely, reducing the search
space to the (typically small) set of nonlinear parameters.

We optimize the nonlinear parameters via Levenberg--Marquardt, minimizing
the reduced objective
\begin{equation}
\min_{\theta_{\rm nl}} \| \vect{y} - \mat{\Phi}(\theta_{\rm nl}) \hat{\vect{c}}(\theta_{\rm nl}) \|^2 + \mu \|\hat{\vect{c}}(\theta_{\rm nl})\|^2 \, ,
\end{equation}
where the gradient through $\hat{\vect{c}}$ is computed analytically via the
implicit function theorem. Thanks to the VarPro decomposition, the optimizer
sees only the nonlinear parameters, with the linear coefficients determined
in closed form at each step, optimal for the current $\theta_{\rm nl}$ by
construction.

\subsection{Template Families}

We currently provide two template families that already cover a useful slice
of nonlinear DE structure.  The first covers power laws $u^p$, $x^p$, and
$(x\cdot u)^p$ with a learnable exponent $p \in [-5, 5]$.  The second covers
exponentials $\exp(kx)$ and $\exp(kx)\cdot u^p$ with a learnable rate
$k \in [-10, 10]$.  Each template is initialized from a data-driven
heuristic (log-log regression for power laws, log-linear for exponentials),
with exponents snapped to nearby integers when close.

Even this restricted template family is enough to move the pipeline
meaningfully beyond a fixed linear library, because exponents or rates
are learned continuous parameters rather than pre-tabulated discrete
choices.

\emph{Support minimization.}  After the template search, we apply
\emph{greedy term removal} to achieve parsimony. For each term in the
discovered equation, we tentatively remove it, refit the remaining
coefficients, and accept the removal if the RMS error increases by less than
a tolerance factor (default $1.1\times$). We iterate this procedure until no
further term can be removed. This post-processing often eliminates spurious
terms that were marginally significant during STLSQ but contribute little to
predictive accuracy.

\emph{Multi-dataset parameter sharing.}  For multi-dataset ordinary
differential equation (ODE) discovery, VarPro supports two modes. In
Phase~1, we perform per-dataset linear refinement, where each dataset
receives independent coefficients $\vect{c}^{(d)}$ while sharing the term
support. In Phase~2, the nonlinear parameters $\theta_{\rm nl}$ (e.g.,
exponents) are shared across all datasets, with per-dataset linear
coefficients $\vect{c}^{(d)}$ still allowed to vary. This enables discovery
of ``universal'' physical laws (same functional form) while accommodating
experiment-specific parameters (different coefficients).

\section{Discovering and Exploiting the Symmetries of the Recovered Law}
\label{sec:symmetry}

The two engines of \S\ref{sec:ode} recover the \emph{form} of a governing
equation. A differential equation also carries a \emph{geometry}, namely its Lie point
symmetries, the continuous transformations of $(x,u)$ that map solutions to
solutions. That geometry can be recovered from the same trajectory data and fed
back into discovery. It lets us (i)~arbitrate between candidates that are
indistinguishable under forward rollout, (ii)~manufacture compositional library
atoms that no fixed dictionary contains, (iii)~integrate a recovered
second-order law to closed form, and (iv), in phase space, derive conserved
charges and the structural coefficients they imply. Throughout, the symmetry is
\emph{discovered from data} and used to amplify or explain the \emph{discovered}
equation, never imposed as a prior. In this it is the reverse of classical
Lie analysis, which starts from a known equation, and complementary to
symmetry-informed sparse regression, which imposes a known invariance to prune a
fixed library. Crucially, none of these levels replaces the discovery engines of
\S\ref{sec:ode}.  Each hands them a candidate they could not otherwise
reach.

\subsection{Determining Certificates for a Discovered Law}

Write a recovered first-order law as the residual $F(x,u,u_x)=u_x-f(x,u)=0$. A
vector field $V=\xi(x,u)\,\partial_x+\eta(x,u)\,\partial_u$ is an infinitesimal
symmetry when its first prolongation annihilates $F$ on shell,
\begin{equation}
V^{(1)}F\big|_{F=0}=0,\qquad
V^{(1)}=V+\big(D_x\eta-u_x\,D_x\xi\big)\,\partial_{u_x},
\label{eq:prolongation}
\end{equation}
where $D_x$ is the total derivative. Restricting $\xi,\eta$ to the affine
family (the translations $\partial_x,\partial_u$, the scalings
$x\partial_x,u\partial_u$, the shears $u\partial_x,x\partial_u$, and their
integer combinations) turns Eq.~\eqref{eq:prolongation} into a linear
determining system whose nullspace, computed by SVD and snapped to integer
generators, is the candidate's certified algebra, within the
searched generator family and the numerical tolerance. The certificate is
\emph{intrinsic}.  The on-shell substitution $u_x=f(x,u)$ uses the
candidate's own right-hand side, so the admitted algebra reflects the
equation's structure and not how well it fits the data. In
the test of \S\ref{sec:operator-factorized-de-compositional}
the alias fits certify no data-supported generator in the searched family,
whereas every true law certifies at least one. The same linear determining construction extends
to quadratic point generators ($\xi,\eta$ of degree $\le2$), reaching the
projective and special-conformal symmetries the affine family cannot
express.

Each certified generator is then tested against the trajectory ensemble through a
flow test. We advance the data a small amount along the generator,
$z\mapsto\exp(\epsilon V)\,z$, and ask whether the flowed points remain
consistent with the same law.  
%
%
The test consults the data but never the
ground-truth labels. The output is, per candidate, a certified Lie algebra
together with a data-support verdict, exactly the information needed to break
ties that forward rollout cannot.

\subsection{Symmetry-Adapted Reduction as a Proposal Amplifier}
\label{sec:symmetry-reduction}

A data-supported generator carries closed-form canonical coordinates $(r,s)$
that straighten its flow, so that $V=\partial_s$. In these coordinates the
reduced equation is univariate, $\dd s/\dd r=G(r)$, and we fit $G$ with the
same mapping families the operator-factorized engine uses: affine, shifted
power with a free exponent, exponential, logarithm, and rational. Pulling
the fitted $G$ back to $(x,u)$ yields a single symbolic term.

That term is handed to the discovery engines in one of three ways: added to
the sparse-library dictionary as an order-one row, supplied as a carrier row
for a second-order fit, or seeded directly into the operator-factorized
engine as a whole-law proposal. In every case the symmetry \emph{amplifies}
rather than replaces the engine.  It supplies a compositional atom that a
fixed dictionary cannot enumerate, one carrying a continuous exponent, a
nested composition, or a shifted-rational form, and lets the ordinary sparse
selection or operator search do the rest. On the first-order test laws of
\S\ref{sec:operator-factorized-de-compositional} the discovered reduction
row is exactly the atom the fixed library lacked, and the sparse fit then
selects it directly.

\subsection{Closed-Form Integration via a Discovered Solvable Algebra}

For a second-order law the same reduction can be applied twice. A data-discovered
scaling $u\partial_u$ reduces the equation to a first-order Riccati whose
equilibria are the characteristic roots when the Riccati is \emph{autonomous} (a
constant-coefficient linear equation) or the indicial roots when it is
\emph{scale-invariant} (an equidimensional Cauchy--Euler equation).  Together with
a second generator these close a two-dimensional solvable algebra, verified
through its Lie bracket, that carries the equation to quadrature in its three
regimes (two real roots, a complex pair, or a repeated root). For instance, from trajectory
data alone the cascade recovers that $u''=-1.69\,u$ integrates to
$u=A\cos(1.3x)+B\sin(1.3x)$ and that the equidimensional $x^2u''+2xu'-6u=0$ has
the power-law solutions $x^{2}$ and $x^{-3}$ (indicial roots $2,-3$), with the
analogous exponential and log-oscillatory forms elsewhere. The cascade is
deliberately conservative. When the reduced Riccati retains explicit
$r$-dependence, as for the Bessel and Lane--Emden equations, whose useful algebra is only
one-dimensional, it reports that no closed form was found rather than forcing a
spurious one.

\subsection{Symplectic Noether Reduction}
\label{sec:symmetry-noether}

For phase-space data the construction takes its symplectic form. We scan
affine generators $\delta z=\mat M z+\vect c$ of the phase space $z=(q,p)$
(rotations, translations, and the canonical dilation), with $\mat M$ infinitesimally 
symplectic ($\mat M^{\!\top}\mat J+\mat J\mat M=0$), and form the Noether momentum
map for each one, which for such a field has the closed
form
\begin{equation}
G(z)=\tfrac{1}{2}\,z^{\!\top}\!\big(-\mat J\mat M\big)z+\big(-\mat J\vect c\big)^{\!\top}z,
\label{eq:momentum-map}
\end{equation}
with $\mat J$ the symplectic matrix ($\mat J^{\!\top}=-\mat J$,
$\mat J^{2}=-\mat I$). Generators whose charge $G$ is conserved along the
trajectories are the system's Noether symmetries, and their charges are the
conserved quantities: linear momentum for translations, angular momentum for
rotations, the virial $q\!\cdot\!p$ for the dilation. Reducing by a
conserved charge lowers the effective dimension and can \emph{derive} a
structural coefficient instead of fitting it.  For a central force the
rotational charge $\mathbf L=\mathbf r\times\mathbf p$ fixes the areal
constant $\ell=|\mathbf L|$ and forces the centrifugal coefficient
$k=\ell^{2}$ (the quadratic Casimir of the recovered rotation algebra), as
demonstrated on real asteroid ephemerides in
\S\ref{sec:kepler-showcase}. For autonomous systems the same construction
extends to noncanonical Poisson geometry, discovering a Jacobi-certified
invariant bivector $\Pi(z)$ (Lie--Poisson when linear) and its Casimir
invariants, rather than presupposing a canonical $(q,p)$ split.
That extension is implemented in the released code but will be
explored in a future contribution.

\section{Benchmarks and Results}
\label{sec:benchmarks}

We evaluate NestyNet-DE on a benchmark suite that stresses differential
equation discovery under held-out generalization.  To our knowledge,
no existing framework has been benchmarked on a collection
that combines real scalar, complex-valued, and coupled vector laws
with held-out trajectory validation.  The closest existing suite, ODEBench 
\citep{dAscoli2024ODEFormer}, collects 63 autonomous first-order systems 
of one to four dimensions, with no units metadata, forcing, explicit dependence 
on the independent variable, or complex-valued and vector-calculus laws
(about ten of its elementary laws have counterparts in our scalar table).
Our suite contains 83 cases.  Of these,
57 are real-valued scalar ordinary differential equations
covering first- and second-order laws drawn from mechanics, circuits, quantum
mechanics, astrophysics, and nonlinear dynamics.  The remaining
26 are complex-valued equations and coupled systems (25 of them
differential and one an algebraic frequency-domain relation included to
exercise the order-zero path),
spanning Schr\"odinger-type models, nonlinear optics, oscillators, and
condensed-matter examples.  The scalar and complex cases differ in their sparse first line.
The scalar cases use the standard candidate library with its
metadata-gated inverse-coordinate extension (\S\ref{sec:ode}),
whereas the complex cases use class-library STLSQ, in which the candidate set
is specialized to each problem's class (\S\ref{sec:complex-laws}).
Throughout the benchmark section we report
PASS/PARTIAL/FAIL.  For the scalar cases the criterion is held-out normalized
root-mean-square error (NRMSE) of the rolled-out law, with thresholds of $10^{-2}$ and
$5\times10^{-2}$, and for the complex cases it is structural agreement with the
answer key (\S\ref{sec:complex-laws}).  Every mark is scored on two
sealed test trajectories that are never used for fitting
or candidate arbitration.  
%
%
Every mark is reported on two sealed trajectories that are never used for fitting
or candidate arbitration.  
%

Each candidate library term and its derivatives are evaluated in closed form
from a trained NestyNet surrogate, so the reported recoveries reflect the
full pipeline of surrogate fitting followed by sparse discovery and
refinement rather than an idealized oracle.  These cases probe the
structural challenges of DE discovery (coupling, complex amplitudes, vector
calculus, and operator factorization).

\subsection{Scalar Differential-Equation Benchmark}

The scalar benchmark is intentionally uniform.  Each problem is represented
as a single real-valued ODE, training data consist of multiple trajectories
with varied initial conditions, and the discovered law is accepted only if
it generalizes under forward simulation on held-out trajectories.  This
setup tests order selection, damping and forcing terms, inverse-coordinate
singularities, and nonlinear self-interactions within one consistent
evaluation protocol.

The fixed-library STLSQ first line already passes a broad cross-section of
the scalar cases, including trigonometric restoring forces, polynomial
nonlinearities, linear damping, and several quantum-mechanical examples with
nontrivial coordinate dependence.  The harder structures, those a fixed
library cannot express, are recovered instead by the structural second line,
the operator-factorized route with the factorized search of Paper~III
available in the same rescue lane: nonlinear self-damping and parametric
forcing, singular inverse-coordinate terms intertwined with derivatives,
and, through the factorized-search lane, half-order kinetics.  The benchmark
is not meant to show that linear regression can rediscover linear response
laws, but to measure how far derivative-accurate discovery reaches into
genuinely nonlinear dynamics, and the structural second line is what carries
it furthest.  Table~\ref{tab:de-benchmark} (Appendix~\ref{app:scalar-table})
summarizes the full benchmark, reporting the scalar cases under two
pipelines, the STLSQ first line alone (``STLSQ'') and the hybrid that adds
the operator-factorized search as a second line (``+OFS'').

The profile of the failures is informative.  STLSQ alone
already achieves PASS on 39 of 57 cases.  Its 18 non-passes cluster around cases
that are awkward for a fixed library: inverse-coordinate factors intertwined
with derivatives or nonlinearities (radial inflow, exact separable,
Lane--Emden), nonlinear self-interaction and parametric forcing (Van der Pol,
Mathieu), and nonlinear denominators involving derivatives (relativistic
oscillator).  This is exactly the part of the benchmark that motivates
structural search, and the +OFS column shows the payoff.  Adding the
structural second line
(\S\ref{sec:operator-factorized-de-compositional}) lifts 13 of these
18 cases to PASS, ten through the operator-factorized route and three
(radial inflow, half-order kinetics, and the exact separable law) through its
factorized-search lane.  
%
%
The driven harmonic oscillator is the instructive lift: its
additive forcing is recovered by a typed lane seeded with periodogram
frequency hints, and a variable-projection polish of the frequency
during the joint refit is what carries the rollout across the PASS
threshold, since over many forcing periods even a percent of frequency
error dephases into an order-of-magnitude residual.

\subsection{Complex-Valued Differential-Equation Benchmark}

The complex benchmark extends the same philosophy to
complex-valued systems by rewriting each problem as a coupled real
system before discovery.  The complex cases deliberately mix multi-trajectory
ODEs with analytic PDE examples,
allowing the same discovery machinery to be exercised on quantum-mechanical
models, nonlinear wave equations, driven oscillators, and condensed-matter
systems without changes to the core pipeline.  
The complex cases are scored structurally, with a pass when every true term is 
recovered within tolerance and no spurious term survives, a partial on a 
coefficient or spurious-term violation, and with a fail when a true term is missing.
Unlike the scalar cases, whose first
line draws candidates from the standard library of
\S\ref{sec:ode}, the complex cases use class-library STLSQ
(\S\ref{sec:complex-laws}).  The candidate set is specialized to each
problem's class (component count, spatial axes, and order), which supplies the
modulus and product structure that complex laws require.

These cases are even more overtly nonlinear than the scalar ones.  The
successful set includes cubic self-interaction, cross-phase modulation,
conjugate coupling, bifurcation normal forms, superconducting gap dynamics,
Josephson nonlinearities, and coupled condensate systems.  These are
nonlinear systems in any physically meaningful sense, even though the outer
regression blocks remain linear in their top-level coefficients.
The complex cases are included in the same combined benchmark table
(Table~\ref{tab:complex-benchmark}, Appendix~\ref{app:complex-table}).

No case fails outright.  Three cases are partial passes.  The hydrogen
radial Schr\"odinger equation and the complex Klein--Gordon field recover
the correct operator structure but retain a single spurious term (a
residual field term and a spurious second-derivative term, respectively).
The time-independent Schr\"odinger eigenvalue problem recovers the
correct support but with aliased coefficients, since its single-mode
eigenfunction data leave the candidate basis nearly collinear.

\subsection{A Controlled Test of Structural Recovery}
\label{sec:operator-factorized-de-compositional}
\label{sec:benchmark-symmetry}

To probe recovery of structure a fixed library cannot express, we hand-build four
first-order laws: three with a coordinate-dependent coefficient,
$u_{x_0}+a(x_0)\,u=0$ for $a(x_0)=1/(1+x_0)$, $1/(1+x_0)^2$, and $\log(1+x_0)$,
and one with a state nonlinearity, $u_{x_0}+\exp(u)=0$, each scored by held-out
rollout.  With the sparse-library line disabled, the operator-factorized route
recovers all four (mean NRMSE $<10^{-2}$), whereas standard STLSQ fails
outright on the two inverse-coordinate laws, whose $u/(1+x_0)$ and $u/(1+x_0)^2$
atoms are absent, and passes only by fitting a local alias on the logarithmic and
exponential laws.  This is the mechanism behind the $+$OFS gains of
Table~\ref{tab:de-benchmark}.

A rollout PASS alone cannot separate a genuine recovery from such an alias, but
the determining certificate of \S\ref{sec:symmetry} can.  For the logarithmic
law, the true equation certifies a data-supported scaling symmetry
($u\partial_u$, at relative RMS $9\times10^{-10}$) while its alias fits certify
nothing.  The same reduction (\S\ref{sec:symmetry-reduction}) then manufactures
the missing atoms $u/(1+x_0)$ and $u/(1+x_0)^2$, which, added as a single column to
the standard library, give exact recovery at NRMSE $\sim\!10^{-8}$.  The
construction likewise recovers the half-order law $u'=-0.6\,\sqrt{u}$ with a free
exponent fitted to $0.500$.  The operator-factorized route supplies the missing
representational step, and the discovered symmetry both certifies it and hands it
to sparse selection.

\subsection{Maxwell Vector-PDE Case Study}

We close the synthetic benchmark with Maxwell's equations, the iconic coupled
vector field laws.  A vector-calculus library curated for electromagnetism
invites an obvious objection that the regression merely echoes back the
operators it was handed.  We therefore broaden the candidate menu beyond
the true terms and ask whether the sparse regression selects Maxwell's
structure while rejecting plausible impostors.  For each component
equation the library is
\begin{equation}
\Phi = \bigl\{\, \nabla\times\mathbf{E},\ \nabla\times\mathbf{B},\
\nabla^{2}\mathbf{E},\ \nabla^{2}\mathbf{B},\ \mathbf{E},\ \mathbf{B}\
(,\ \mathbf{J})\,\bigr\},
\end{equation}
with coefficients tied across the three spatial components.  The second-order
Laplacians $\nabla^{2}\mathbf{E},\nabla^{2}\mathbf{B}$ are genuine decoys of the
same differential order as the diffusive term one would add for a conductor.
To isolate the selection question from surrogate accuracy, we first
supply exact tabulated field values and derivatives, obtaining the analytic
spatial Hessian required by the $\nabla^{2}$ terms from spectral differentiation.

\begin{figure*}
\centering
\includegraphics[width=\textwidth]{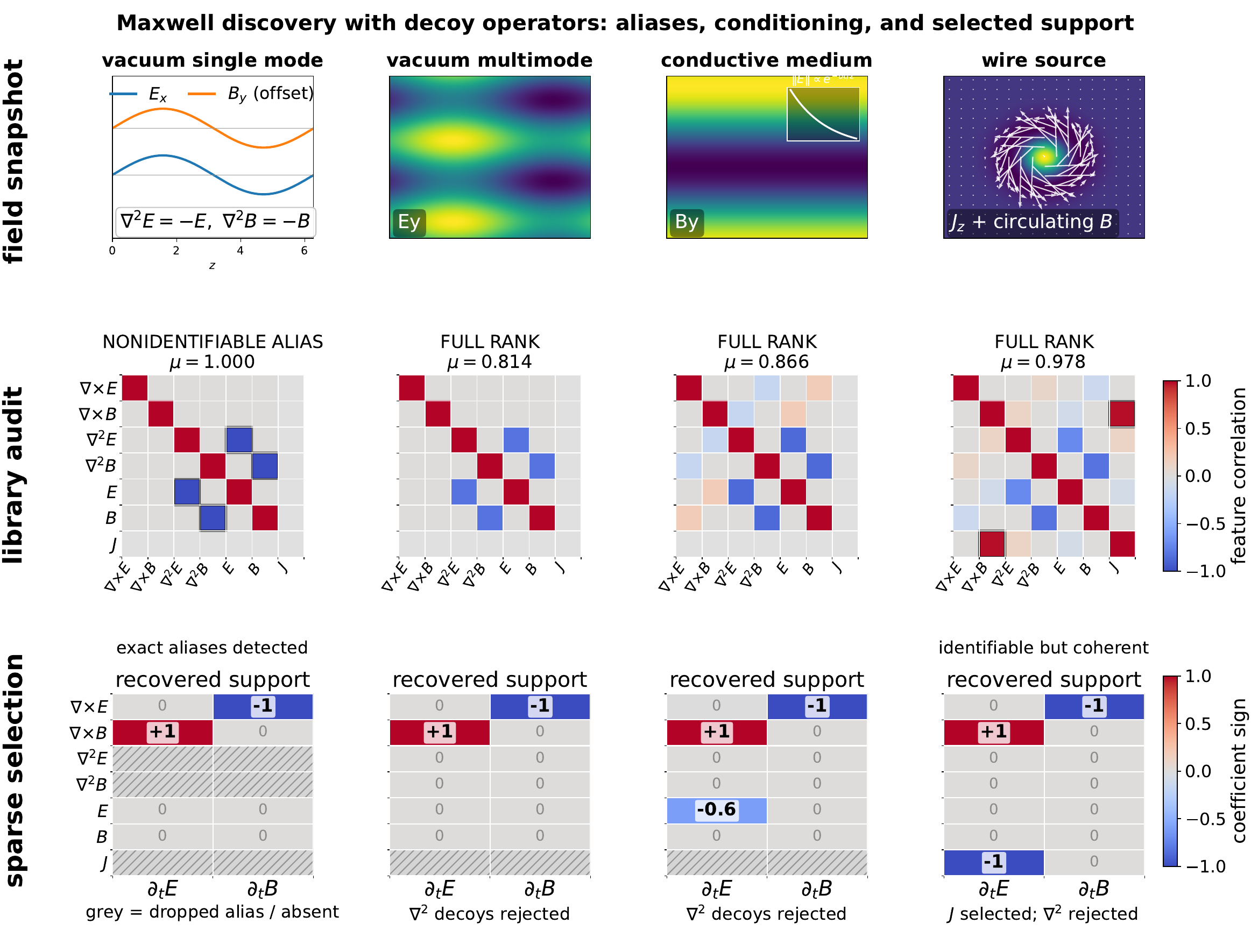}
\caption{Discovery of the Maxwell equations from data in four physical scenarios, using a broad library.
\emph{Top:} a representative field snapshot.  The single-mode wave
is a one-dimensional profile, the others are slices, with the conductor's
$e^{-\sigma t/2}$ envelope inset.  \emph{Middle:} the normalized-column Gram
matrix of the candidate library, annotated with the largest off-diagonal
correlation in magnitude, $\mu$.  The single-mode vacuum shows exact aliases
$\nabla^{2}\mathbf{E}\equiv-\mathbf{E}$, $\nabla^{2}\mathbf{B}\equiv-\mathbf{B}$
(boxed) and a rank-deficient design, whereas the wire is identifiable but highly
coherent (boxed $\nabla\times\mathbf{B}$--$\mathbf{J}$ cell).  \emph{Bottom:} the
recovered coefficients in the $\partial_{t}\mathbf{E}$ and
$\partial_{t}\mathbf{B}$ equations.  The curls (with $\sigma\mathbf{E}$ for the
conductor and $\mathbf{J}$ for the wire) are selected while every Laplacian and
spurious algebraic decoy is driven to zero (shown as ``$0$'').  Hatching marks the alias columns dropped in favor of their
lowest-order representative.}
\label{fig:maxwell-decoy}
\end{figure*}

Figure~\ref{fig:maxwell-decoy} summarizes four
excitations of the electromagnetic field.  In the three
identifiable cases the broad library recovers exactly the Maxwell operators
$\partial_{t}\mathbf{E}=\nabla\times\mathbf{B}$ (vacuum and multi-mode), with an
added $-\sigma\mathbf{E}$ in the conductor ($\sigma=0.6$, recovered to
$2\times10^{-4}$) and $-\mathbf{J}$ for the wire source, and
$\partial_{t}\mathbf{B}=-\nabla\times\mathbf{E}$ throughout, while the
$\nabla^{2}$ and spurious algebraic terms are driven to zero, leaving a residual at machine
precision ($\lesssim 10^{-14}$).

The fourth excitation, a single linearly polarized plane wave, is deliberately
reported as a degeneracy check rather than a recovery.  Each active field
component is a Laplacian eigenfunction, so $\nabla^{2}\mathbf{E}=-k^{2}\mathbf{E}$
and $\nabla^{2}\mathbf{B}=-k^{2}\mathbf{B}$ exactly.  The library columns
are then exact aliases and the design is rank-deficient.  A pre-regression
conditioning audit (the normalized-column Gram matrix, its numerical rank,
and per-term variance-inflation factors) detects this, and a rank-aware solve
reports the support as correct only \emph{modulo the alias class}
$\mathbf{E}\sim-\nabla^{2}\mathbf{E}$, retaining the lowest-order representative
by a declared convention rather than by physical preference.  Identifiability is
then restored purely by excitation.  A superposition of vacuum plane waves with
distinct $|\mathbf{k}|$ (the multi-mode row) is an exact source-free solution
whose Laplacian is no longer proportional to the field, and the broad library
again selects only the curls.  The wire case is identifiable but highly coherent
(peak inter-column correlation $0.98$, selected-support variance inflation
$\approx 23$), reflecting the physical correlation between $\nabla\times\mathbf{B}$
and its source $\mathbf{J}$, so its noise sensitivity is intrinsic to the problem's conditioning.

Finally, we replace the exact derivatives with a NestyNet segmented surrogate
that supplies its own analytic derivatives, trained by Sobolev
continuation (values first, then a geometrically increasing derivative-channel
weight, with a deterministic restart to clear a residual $C^{1}$-suboptimal basin
in one component).  
%

Although the surrogate does not reach oracle-level derivative accuracy, it
reaches operational discovery accuracy.  On the conductive case at
$10^{-6}$ relative noise it recovers the full Maxwell support
($\nabla\times\mathbf{B}$, $\nabla\times\mathbf{E}$, and $\sigma\mathbf{E}$, all
within tolerance) while rejecting both Laplacian decoys (the residual
false-discovery rate of $0.25$ in Table~\ref{tab:maxwell-surrogate} reflects
a single sub-percent leak of the field $\mathbf{B}$, not a decoy), whereas local
finite-difference and unregularized spectral derivatives of the same noisy field
activate the decoys at every noise level.  A noise ladder
(Table~\ref{tab:maxwell-surrogate}), run with the dimensional prior
deliberately disabled so that the stress test stays units-blind, shows this
clean rejection holds to $10^{-6}$ relative noise and collapses by $10^{-5}$, 
so we characterize the units-blind method as achieving operational recovery 
in this noise ladder up to ${\sim}10^{-6}$ relative noise.

\begin{table*}[t]
\centering
\caption{Operator selection versus derivative source on the conductive-medium
Maxwell case, with the broadened library
$\{\mathbf{E},\mathbf{B},\nabla\times\mathbf{E},\nabla\times\mathbf{B},
\nabla^{2}\mathbf{E},\nabla^{2}\mathbf{B}\}$.  The true support is
$\{\nabla\times\mathbf{B},\nabla\times\mathbf{E},\sigma\mathbf{E}\}$ and the two
Laplacians are the decoys.  The true support is recovered in every row
(true-positive rate ${\rm TPR}{=}1$), so the discriminator is whether the second-order decoys are
\emph{rejected}, as measured by the false-discovery rate (FDR).  Classical finite-difference and spectral derivatives
select both decoys at every noise level (discretization/time-difference bias, not
a noise effect).  The \emph{denoised-spectral} row is a direct spectral feature
table built from the low-passed noisy field with the same oracle-time anchor the
surrogate uses.  It rejects both decoys to $10^{-4}$, so the noisy data carry
enough information.  The learned NestyNet Sobolev surrogate rejects the decoys only to
${\sim}10^{-6}$ and breaks by $10^{-5}$, a pipeline limit (the $H^{1}$-trained
surrogate's unsupervised curvature) rather than a data limit.  The final block
repeats the surrogate ladder with the declared-constant unit prior of \S\ref{sec:units:de}
active.  Both Laplacian decoys are then pruned before regression at every noise
level, so the recovered support stays correct across the full ladder while the
coefficient accuracy still degrades with noise.  A single noise realization per level is used.}
\label{tab:maxwell-surrogate}

\begin{tabular}{llcccc}
\hline
Derivative source & Noise & FDR & $\nabla^{2}$ decoys & Coeff.\ err & Residual \\
\hline
Exact (oracle)        & $0$--$10^{-4}$ & $0.00$ & rejected      & $2\times10^{-14}$ & $9\times10^{-15}$ \\
Finite difference     & $0$--$10^{-4}$ & $0.50$ & both selected & $5.4\%$  & $18\%$  \\
Spectral (FFT)        & $0$--$10^{-4}$ & $0.50$ & both selected & $18\%$   & $12\%$  \\
Denoised spectral$^{\dagger}$ & $0$--$10^{-4}$ & $0.00$ & rejected & ${\le}4\times10^{-7}$ & ${\le}2\times10^{-5}$ \\
NestyNet Sobolev surrogate & $0$       & $0.25$ & rejected      & $0.11\%$ & $1.4\%$ \\
                      & $10^{-6}$ & $0.25$ & rejected      & $2.7\%$  & $10\%$  \\
                      & $10^{-5}$ & $0.50$ & both selected & $32\%$   & $89\%$  \\
                      & $10^{-4}$ & $0.50$ & both selected & $40\%$   & $47\%$  \\
NestyNet surrogate $+$ unit prior & $0$ & $0.25$ & pruned (units) & $0.11\%$ & $1.4\%$ \\
                      & $10^{-6}$ & $0.25$ & pruned (units) & $2.7\%$  & $10\%$  \\
                      & $10^{-5}$ & $0.25$ & pruned (units) & $17\%$   & $141\%$ \\
                      & $10^{-4}$ & $0.25$ & pruned (units) & $40\%$   & $47\%$  \\
\hline
\multicolumn{6}{l}{\footnotesize $^{\dagger}$ direct spectral feature table from the
low-passed noisy field, oracle-time anchor (a data upper bound, not a deployable method).}\\
\end{tabular}
\end{table*}

This ${\sim}10^{-6}$ ceiling is a property of the learned surrogate, not of the
data.  With the same exact time-derivative anchor as the surrogate, a direct spectral feature table from
the same noisy field (Table~\ref{tab:maxwell-surrogate}, ``denoised spectral'')
rejects both decoys to $10^{-4}$.  The bottleneck is the surrogate's own
curvature, trained on values and first derivatives ($H^{1}$) but evaluated through
its unsupervised second derivatives ($H^{2}$), so lifting the ceiling is a matter
of $H^{2}$-supervised or band-limited training rather than better first-order gate
placement.  The clean end of the ladder is not an idealization. Precision-physics 
data routinely reach relative accuracies of $10^{-6}$ and beyond, and a 
second-order optimizer is what would let law discovery exploit them, since it 
can converge toward machine precision where first-order training stalls.

The ceiling also applies only to the units-blind stress test.  In the
pipeline the declared-constant unit prior of \S\ref{sec:units:de} is active
(a coefficient is admissible only if its dimension matches a declared constant),
and under the declared units both Laplacian columns require an undeclared
diffusivity coefficient, so they are pruned before any regression at every
noise level\footnote{The declared constants are matched atomically.  The
engine compares each required coefficient dimension against the declared
constants themselves and never forms their products or ratios, so declaring
$c^{2}$ and $\sigma$ does not implicitly admit a diffusivity $c^{2}/\sigma$.}.
All true terms then remain selected and both decoys pruned across the full ladder to $10^{-4}$
relative noise (Table~\ref{tab:maxwell-surrogate}, final block), with the
field-$\mathbf{B}$ term the only spurious selection throughout
(dimension-admissible in the induction equation as a magnetic damping term,
which is why the prior cannot exclude it), sub-percent at low noise and
growing with the overall coefficient error at high noise.  The dimensional
prior thus preserves the true support well past the units-blind
ceiling, and the $H^{2}$ route above remains the lever for coefficient
precision.

In the Maxwell experiments we use the standard vector-calculus operators as the
physically interpretable library, because the present goal is to test
vector-PDE discovery and derivative accuracy.  A natural extension is to replace
these operators by primitive derivative tensors and allow the factorized search
to recover the invariant contractions corresponding to curl and divergence.

\subsection{Recovering the Reduced Kepler Hierarchy from Real Ephemerides}
\label{sec:kepler-showcase}

To complement the above synthetic tests with a real-data theory-discovery
capstone, we show that from 30 years of daily trajectories obtained from
NASA's Jet Propulsion Laboratory (JPL) HORIZONS ephemeris
service~\citep{GiorginiHORIZONS} for 308 main-belt asteroids, the method
rediscovers the reduced Kepler hierarchy.

\begin{figure*}[t]
    \centering
    \includegraphics[width=0.78\textwidth]{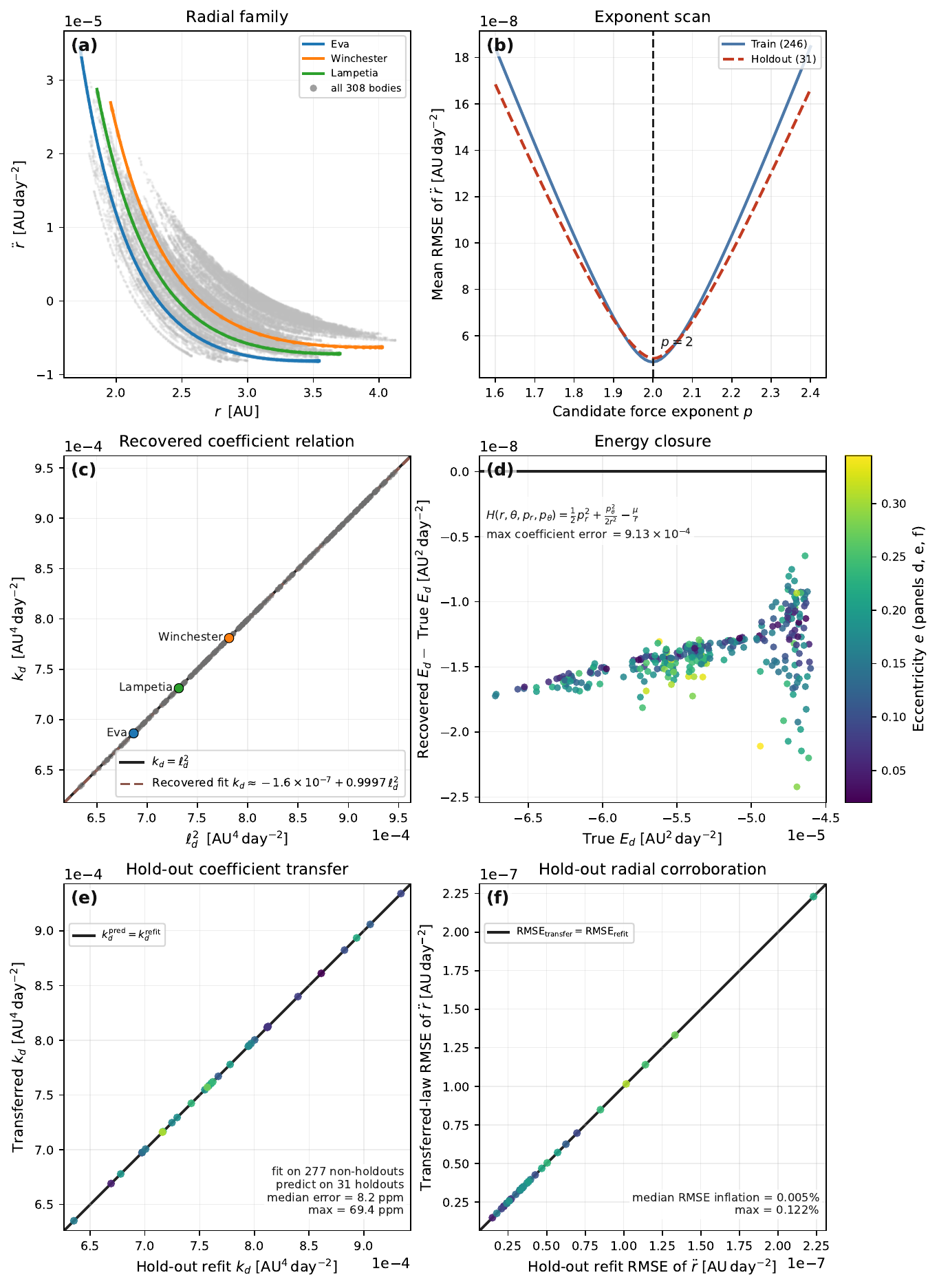}
    \caption{Middle and upper rungs of the recovered hierarchy on the sample of 
    asteroids. (a) Radial accelerations of all $308$ bodies (gray) with the three 
    highest-leverage orbits highlighted. Each colored curve is the recovered radial family 
    $\ddot r = k_d/r^3 - \mu_\odot/r^2$ evaluated with that body's $k_d$.  (b) The $308$ 
    bodies are split deterministically into $246$ train, $31$ 
    validation, and $31$ hold-out objects, and under that split the inverse-square law 
    is still selected on hold-out bodies. (c) The independently recovered centrifugal 
    coefficients obey the recovered coefficient relation, which is nearly indistinguishable 
    from the Kepler identity $k_d=\ell_d^2$. (d) The reduced-energy post-pass closes 
    the loop in residual form, the energy errors remaining small across the ensemble, 
    while the broader spread toward the less-bound, outer-belt end marks where the 
    real ephemerides depart most from ideal reduced two-body closure. Panels (e) 
    and (f) give the stricter cross-orbit corroboration on the same deterministic split, 
    with the coefficient relation $k_d = a + b\,\ell_d^2$ fit on the $277=246+31$ 
    non-hold-out bodies and transferred to the same $31$ unseen hold-out bodies without 
    refitting $k_d$. (e) The predicted hold-out $k_d$ agree closely with independently 
    refit hold-out values. (f) On those hold-out bodies the transferred radial law yields nearly 
    the same $\ddot r$ RMSE as the best per-orbit refit using the same shared $\mu_\odot$.}
    \label{fig:kepler-radial-manifold}
    \label{fig:kepler-holdout-corroboration}
\end{figure*}

The sample is drawn from JPL small-body ephemerides cross-matched to
curated asteroid masses from the SsODNet `ssoBFT' best-estimate
table~\citep{Berthier2023} (numbered main-belt asteroids with semi-major axis
$2.1<a<3.3$\,AU, perihelion $q>1.6$\,AU, an observation arc 
$>15\,000$\,days with at least $200$ observations, and an SsODNet mass above
$10^{17}$\,kg).

We deliberately use the weathered trajectories themselves, the heliocentric
HORIZONS state vectors from 1980--2009 at 1-day cadence.  These are
JPL's best-fit dynamical ephemerides, not raw observations, so they carry
the full Solar-System perturbations rather than idealized two-body motion.
Any reduced-Kepler structure that emerges is therefore recovered as the
dominant skeleton hidden inside perturbed ephemerides rather than imposed
by construction.

The radial accelerations that feed the ladder are analytic first derivatives
of per-body NestyNet surrogates fitted to the ephemeris velocity channels on
a data-discovered cylinder chart (a circle at the dominant measured
frequency times a slow drift axis that absorbs the planetary weathering), so
no finite differencing enters this primary inverse-square discovery pipeline.
Fitting the position channels as well and scoring their analytic derivatives 
against the velocity data gives a median relative error of $7.9\times10^{-7}$
as an order-matched diagnostic of the acceleration error at 1-day cadence.

Operationally, the NestyNet-DE pipeline is asked to recover shared structure
across the ensemble. The analysis sequence forms a discovery
ladder. First, the angular dynamics collapse onto Kepler's areal law
\begin{equation}
\dot{\theta} = \frac{\ell_d}{r^2} \, ,
\end{equation}
with one orbit-specific areal constant $\ell_d$. Second, the radial dynamics are recovered as
\begin{equation}
\ddot r = \frac{k_d}{r^3} - \frac{\mu_\odot}{r^2} \, ,
\end{equation}
namely an orbit-specific centrifugal barrier plus a shared inverse-square attraction.
Third, the independently recovered coefficients fall onto the relation
\begin{equation}
k_d = \ell_d^2 \, ,
\end{equation}
showing that the $r^{-3}$ term is the square of the areal-law constant itself.
Fourth, a reduced-energy post-pass recovers
\begin{equation}
E_d = \frac{1}{2}\dot r^2 + \frac{\ell_d^2}{2r^2} - \frac{\mu_\odot}{r} \, .
\end{equation}
Only after these pieces are in place do we assemble the reduced Hamiltonian
\begin{equation}
H(r,\theta,p_r,p_\theta)=\frac{1}{2}p_r^2+\frac{p_\theta^2}{2r^2}-\frac{\mu_\odot}{r} \, .
\end{equation}

The example works because the collapse begins at the very first rung.
Already there, the full $308$-body daily ensemble compresses onto a
one-parameter areal-law family, and the recovered areal-law constants
agree with the orbital values to maximum relative error $2.55\times 10^{-4}$.
Separately, the reduced-energy post-pass recovers the expected energy
coefficients with maximum absolute coefficient error $9.13\times 10^{-4}$.

Figure~\ref{fig:kepler-radial-manifold} contains the decisive middle and
upper steps. The shared solar parameter is recovered as
$\mu_\odot = 2.9573\times 10^{-4}$ in the adopted heliocentric units
(AU$^{3}$\,d$^{-2}$), about $0.06\%$ below the reference. On a deterministic radial-leverage
split ($246$ train, $31$ validation, $31$ hold-out), the exponent
selected on the training bodies is $p=2$, and the validation and untouched
hold-out error curves independently attain their minima there.  The centrifugal coefficients
satisfy $\max_d |k_d-\ell_d^2| = 1.00\times 10^{-6}$. Panel~(d) closes
the loop in residual form.  The energy errors remain small across the
ensemble, and the broadening toward the less-bound end shows the
remaining non-Kepler weathering localizing to the outer belt.

Panels (e) and (f) supply the stricter cross-orbit corroboration on that
same split. Fitting the coefficient relation on the $277$ non-hold-out bodies
and transferring it unchanged to the $31$ hold-out bodies predicts their coefficients
to a median relative error of $8\times10^{-6}$, and the transferred radial
law is essentially indistinguishable from the best per-orbit refit (median
$\ddot r$ RMSE inflation $4\times10^{-5}$). The learned cross-orbit relation
therefore genuinely transfers rather than merely draws a plausible family of curves.

We emphasize that the final Hamiltonian is assembled from recovered
symbolic ingredients rather than discovered as a single expression, and
that we do not begin by postulating it. Even so, the direct family recovery,
the hold-out exponent scan, the cross-orbit corroboration, and the energy
closure together indicate that the pipeline extracts the full reduced-Kepler
hierarchy, including $k_d=\ell_d^2$, from perturbed ephemerides of real
asteroids rather than overfitting one favorable orbit at a time.

The coefficient identity $k_d=\ell_d^2$ need not remain an empirical
cross-orbit fit.  It can be \emph{derived} from a single symmetry that the
data themselves select. Working directly from the Cartesian phase-space
trajectories $(\mathbf r,\dot{\mathbf r})$, without introducing any angle or
polar coordinate, we scan a family of candidate phase-space generators
(spatial rotations, translations, and the canonical dilation
$q\to e^{\epsilon}q$, $p\to e^{-\epsilon}p$), form the Noether
momentum map of each, and retain those whose charge is conserved along
the ephemerides. 
%
%
On the full $308$-body ensemble the three rotations survive, with charge drift 
(the maximum excursion of $G$ relative to $\max_t|\mathbf q||\mathbf p|$)
between $5\times10^{-4}$ and $10^{-3}$, while the translations ($0.07$--$0.39$) 
and the dilation ($0.16$) are rejected.  The small residual drift of
the admitted rotations is itself the planetary-perturbation floor rather than
numerical noise. The admitted charge is the angular momentum
$\mathbf L=\mathbf r\times\mathbf p$, whose conservation fixes the areal
constant $\ell=|\mathbf L|=r^2\dot\theta$.  Eliminating the cyclic angle then
yields $\ddot r=\ell^2/r^3-\mu_\odot/r^2$ directly, so the centrifugal coefficient
$k=\ell^2$ is \emph{forced} by the discovered symmetry rather than fitted.
Measured against this derived identity, the independent Noether reduction
gives $k_d/\ell_d^2$ with median $0.9991$ and maximum deviation
$4.45\times10^{-3}$ across the $308$ bodies, and returns
$\mu_\odot=2.9559\times10^{-4}$, about $0.11\%$ below the reference. 
A single discovered symmetry thus supplies both the
areal law and the centrifugal term, and the reduced Hamiltonian
$H=\tfrac{1}{2} p_r^2+\ell^2/(2r^2)-\mu_\odot/r$ follows without postulating its form.

\section{Conclusions}
\label{sec:conclusions}

We have presented NestyNet-DE, a framework for discovering
scalar, vector, or complex-valued differential
equations from data using the analytic derivatives of a trained neural
surrogate.  Two discovery routes exploit these derivatives.
The first follows SINDy and PDE-FIND, assembling a library of candidate terms
and selecting a parsimonious law by sparse regression.  The second searches
directly over typed operator factorizations, arbitrating among candidates by
forward rollout rather than by selection from a fixed library.

These results show that derivative-accurate, interpretable
discovery reaches the nonlinear, complex-valued, and coupled differential
equations that define much of modern physics, the regimes where
numerical-derivative and scalar-only methods break down.  The benchmark
spans many of the flagship equations of physics, from the nonlinear pendulum and Duffing
oscillator to the Gross--Pitaevskii, nonlinear Schr\"odinger, and
Ginzburg--Landau equations and Maxwell's system, and to our knowledge no
previous framework
recovers governing equations across this combination of real,
complex-valued, and vector laws directly from data while keeping
the recovered law explicit.  These settings are not merely scalar SINDy repeated
several times.  They demand consistent cross-channel derivatives, shared
supports, tied coefficients, and structured operator libraries.  Our results establish analytic surrogates,
structured sparse discovery, and post-selection nonlinear refinement as a
practical engine for interpretable equation discovery in physics.

The reduced-Kepler showcase provides a vivid illustration of the algorithm. On a
large ensemble of $308$ main-belt asteroid ephemerides,
using the weathered HORIZONS trajectories, the pipeline recovers the reduced hierarchy from
areal law through inverse-square radial flow to the natural Hamiltonian.
That an exact two-body hierarchy emerges from hundreds of independent,
perturbed orbits at once (not from a single curated trajectory) is a
uniquely stringent test of data-driven theory recovery, and the method passes
it.

Differential-equation discovery is not ordinary SR
with the highest derivative relabeled as the target variable.  Trajectory data
constrain the law only on a thin dynamical manifold in $(x,u,u_x,\ldots)$ space,
where pointwise-equivalent expressions can generate very different vector fields
under rollout.  This is why a DE-native abstraction with dynamical validation,
rather than a pointwise fit, is the right frame.  It also points to a natural
next step of actively choosing the trajectories or experiments that most sharply
separate competing laws, so that discovery is constrained where it would
otherwise be weakly determined.

Beyond a law's form, the framework recovers its geometry, closing the loop.
The Lie point symmetries of a recovered equation, discovered from the same
trajectories, are fed back to arbitrate between candidates that forward rollout
cannot separate (\S\ref{sec:benchmark-symmetry}), to manufacture the
compositional atoms a fixed library lacks, and, for solvable equations, to
integrate the recovered law in closed form. In phase space the construction is
symplectic.  The conserved Noether charges are discovered from the trajectories,
and for the reduced-Kepler problem the rotational symmetry forces the centrifugal
coefficient $k=\ell^2$ rather than fitting it (\S\ref{sec:kepler-showcase}).
In every case the geometry is discovered from the data and tied to the discovered
equation, serving both as an input that amplifies discovery and as the output that
explains and integrates the law.

A direction worth exploring is a complex-native symbolic backend.  Our AST and
analytic-derivative approach should be generalizable to tensor field equations
and, ultimately, to differential-geometric settings in which the operator
library itself depends on metric or connection structure.
Implementing such extensions would bring data-driven
equation discovery closer to the coupled, multi-field, and geometrically
structured laws of particular interest to modern theoretical physics.

\clearpage
\section*{Data Availability}

NestyNet-DE is contained within NestyNet-SR, which is available at
\url{https://github.com/RodrigoIbata/NestyNet_SR}.  The reproducibility package 
for the present work can be found at 
\dataset[doi:10.5281/zenodo.22027498]{https://doi.org/10.5281/zenodo.22027498}.
It contains the exact benchmark inputs, all 308 daily HORIZONS
ephemerides, the corresponding per-body analytic-surrogate acceleration cache
with its velocity-consistency diagnostics, compact summaries for all fifteen protocol steps,
the final figures, provenance records, and file-level checksums.  The archive
records the JPL HORIZONS and SsODNet provenance of the dataset selection.

\begin{acknowledgments}
RI gratefully acknowledges funding in the initial stages of this project from the European Research Council (ERC) under the European Union's Horizon 2020 research and innovation program (grant agreement No. 834148). We gratefully acknowledge the High Performance Computing center of the Universit\'e de Strasbourg for a very generous time allocation and for their support over the development of this project.
\end{acknowledgments}

\software{NestyNet \citep{NestyNet2026a},
PyTorch \citep{Paszke2019},
NumPy \citep{Harris2020},
SciPy \citep{Virtanen2020},
SymPy \citep{Meurer2017},
Matplotlib \citep{Hunter2007}.}

\appendix

\section{Differential-Equation Benchmark Table}
\label{app:scalar-table}\label{app:complex-table}

\startlongtable
\begin{deluxetable*}{@{}llccc@{}}
\tabletypesize{\scriptsize}
\tablecaption{Differential-equation discovery benchmark. The 57 real-valued scalar cases (blank in the $\mathbb{C}$ column) are each scored under two pipelines, first-line STLSQ alone (``STLSQ'') and the hybrid that adds the operator-factorized search as a second-line rescue (``+OFS'').  On the sealed test trajectories STLSQ alone yields 39~\passmark, 9~\partialmark, and 9~\failmark, and adding the OFS second line yields 52~\passmark, 2~\partialmark, and 3~\failmark.  On the validation trajectories, the tallies are 40/8/9 and 55/2/0. The 26 complex-valued cases (marked $\checkmark$ in the $\mathbb{C}$ column) are recovered by class-library STLSQ (\S\ref{sec:complex-laws}), that is, STLSQ over a metadata-driven complex class library specialized to each problem's class rather than a generic dictionary or a per-problem oracle. Their result appears in the STLSQ column, and the operator-factorized second line is not applied to them, so the +OFS entry is left as~\nodata. The complex cases yield 23~\passmark, 3~\partialmark, and no~\failmark, and the equation scoring each is held separately as a ground-truth term list that never enters the discovery library, so the candidate set is class-informed but answer-blind.
%
\label{tab:de-benchmark}\label{tab:complex-benchmark}}
\tablehead{\colhead{Name} & \colhead{Formula} & \colhead{$\mathbb{C}$} & \colhead{STLSQ} & \colhead{+OFS}}
\startdata
\pc{0.32\textwidth}{Radioactive decay} & \pc{0.36\textwidth}{$\displaystyle \frac{d u}{d t} = - \lambda u$} &  & \passmark & \passmark \\
\pc{0.32\textwidth}{Exponential growth} & \pc{0.36\textwidth}{$\displaystyle \frac{d N}{d t} = r N$} &  & \passmark & \passmark \\
\pc{0.32\textwidth}{Logistic growth} & \pc{0.36\textwidth}{$\displaystyle \frac{d N}{d t} = r N \left(1 - \frac{N}{K}\right)$} &  & \passmark & \passmark \\
\pc{0.32\textwidth}{Newton's law of cooling} & \pc{0.36\textwidth}{$\displaystyle \frac{d T}{d t} = - k \left(T - T_{env}\right)$} &  & \passmark & \passmark \\
\pc{0.32\textwidth}{Steady-state heat conduction (1D)} & \pc{0.36\textwidth}{$\displaystyle \frac{\partial T}{\partial x} = -\frac{T}{L}$} &  & \passmark & \passmark \\
\pc{0.32\textwidth}{RC circuit discharge} & \pc{0.36\textwidth}{$\displaystyle \frac{d q}{d t} = -\frac{q}{R C}$} &  & \passmark & \passmark \\
\pc{0.32\textwidth}{RL circuit decay} & \pc{0.36\textwidth}{$\displaystyle \frac{d I}{d t} = -\frac{R}{L} I$} &  & \passmark & \passmark \\
\pc{0.32\textwidth}{RC charging} & \pc{0.36\textwidth}{$\displaystyle \frac{d q}{d t} = \frac{V_{0} - \frac{q}{C}}{R}$} &  & \passmark & \passmark \\
\pc{0.32\textwidth}{Falling with linear drag} & \pc{0.36\textwidth}{$\displaystyle \frac{d v}{d t} = g - \frac{k v}{m}$} &  & \passmark & \passmark \\
\pc{0.32\textwidth}{Falling with quadratic drag} & \pc{0.36\textwidth}{$\displaystyle \frac{d v}{d t} = g - \frac{k v^{2}}{m}$} &  & \passmark & \passmark \\
\pc{0.32\textwidth}{Radial inflow (mass conservation)} & \pc{0.36\textwidth}{$\displaystyle \frac{d v}{d r} = -\frac{v}{r}$} &  & \partialmark & \passmark \\
\pc{0.32\textwidth}{Hydrostatic pressure} & \pc{0.36\textwidth}{$\displaystyle \frac{d p}{d x} = -\rho g$} &  & \passmark & \passmark \\
\pc{0.32\textwidth}{Tunneling wavefunction (evanescent)} & \pc{0.36\textwidth}{$\displaystyle \frac{\partial \psi}{\partial x} = - \kappa \psi$} &  & \passmark & \passmark \\
\pc{0.32\textwidth}{Second-order reaction} & \pc{0.36\textwidth}{$\displaystyle \frac{d c}{d t} = -k c^{2}$} &  & \passmark & \passmark \\
\pc{0.32\textwidth}{Half-order reaction} & \pc{0.36\textwidth}{$\displaystyle \frac{d c}{d t} = - k \sqrt{c}$} &  & \failmark & \passmark \\
\pc{0.32\textwidth}{Simple harmonic oscillator} & \pc{0.36\textwidth}{$\displaystyle \frac{d^2 x}{d t^2} = - \omega^{2} x$} &  & \passmark & \passmark \\
\pc{0.32\textwidth}{Mass-spring system} & \pc{0.36\textwidth}{$\displaystyle \frac{d^2 x}{d t^2} = -\frac{k}{m} x$} &  & \passmark & \passmark \\
\pc{0.32\textwidth}{Simple pendulum (small angle)} & \pc{0.36\textwidth}{$\displaystyle \frac{d^2 \theta}{d t^2} = -\frac{g}{L} \theta$} &  & \passmark & \passmark \\
\pc{0.32\textwidth}{Damped harmonic oscillator} & \pc{0.36\textwidth}{$\displaystyle \frac{d^2 x}{d t^2} = - \omega_{0}^{2} x - 2 \gamma \frac{dx}{dt}$} &  & \passmark & \passmark \\
\pc{0.32\textwidth}{Damped mass-spring} & \pc{0.36\textwidth}{$\displaystyle \frac{d^2 x}{d t^2} = - \frac{k}{m} x - \frac{b}{m} \frac{dx}{dt}$} &  & \passmark & \passmark \\
\pc{0.32\textwidth}{RLC circuit} & \pc{0.36\textwidth}{$\displaystyle \frac{d^2 q}{d t^2} = - \frac{R}{L} \frac{dq}{dt} - \frac{q}{L C}$} &  & \passmark & \passmark \\
\pc{0.32\textwidth}{Free fall} & \pc{0.36\textwidth}{$\displaystyle \frac{d^2 x}{d t^2} = - g$} &  & \passmark & \passmark \\
\pc{0.32\textwidth}{Free fall with linear drag} & \pc{0.36\textwidth}{$\displaystyle \frac{d^2 x}{d t^2} = - g - k \frac{dx}{dt}$} &  & \passmark & \passmark \\
\pc{0.32\textwidth}{Radial orbit (1D collapse)} & \pc{0.36\textwidth}{$\displaystyle \frac{d^2 r}{d t^2} = -\frac{G M}{r^{2}}$} &  & \passmark & \passmark \\
\pc{0.32\textwidth}{Binet orbit equation} & \pc{0.36\textwidth}{$\displaystyle \frac{d^2 u}{d \phi^2} = \frac{G M}{L^{2}} - u$} &  & \passmark & \passmark \\
\pc{0.32\textwidth}{Poisson equation (1D)} & \pc{0.36\textwidth}{$\displaystyle \frac{\partial^2 \phi}{\partial x^2} = -\frac{\rho}{\epsilon_{0}}$} &  & \passmark & \passmark \\
\pc{0.32\textwidth}{Evanescent EM wave} & \pc{0.36\textwidth}{$\displaystyle \frac{\partial^2 E}{\partial x^2} = k^{2} E$} &  & \passmark & \passmark \\
\pc{0.32\textwidth}{Propagating EM wave} & \pc{0.36\textwidth}{$\displaystyle \frac{\partial^2 E}{\partial x^2} = -k^{2} E$} &  & \passmark & \passmark \\
\pc{0.32\textwidth}{Lane-Emden equation} & \pc{0.36\textwidth}{$\displaystyle \frac{d^2 \theta}{d \xi^2} = - \frac{2}{\xi} \frac{d\theta}{d\xi} - \theta^{n}$} &  & \partialmark & \failmark \\
\pc{0.32\textwidth}{Lane-Emden n=1 (polytrope)} & \pc{0.36\textwidth}{$\displaystyle \frac{d^2 \theta}{d \xi^2} = - \frac{2}{\xi} \frac{d\theta}{d\xi} - \theta$} &  & \partialmark & \passmark \\
\pc{0.32\textwidth}{Lane-Emden n=3 (polytrope)} & \pc{0.36\textwidth}{$\displaystyle \frac{d^2 \theta}{d \xi^2} = - \frac{2}{\xi} \frac{d\theta}{d\xi} - \theta^{3}$} &  & \partialmark & \passmark \\
\pc{0.32\textwidth}{Isothermal sphere} & \pc{0.36\textwidth}{$\displaystyle \frac{d^2 \theta}{d \xi^2} = - \frac{2}{\xi} \frac{d\theta}{d\xi} - e^{-\theta}$} &  & \failmark & \passmark \\
\pc{0.32\textwidth}{Bessel equation} & \pc{0.36\textwidth}{$\displaystyle \frac{d^2 y}{d x^2} = \left(\frac{\nu^{2}}{x^{2}} - 1\right) y - \frac{1}{x} \frac{dy}{dx}$} &  & \passmark & \passmark \\
\pc{0.32\textwidth}{Bessel J\_0} & \pc{0.36\textwidth}{$\displaystyle \frac{d^2 y}{d x^2} = - y - \frac{1}{x} \frac{dy}{dx}$} &  & \partialmark & \passmark \\
\pc{0.32\textwidth}{Time-indep Schrodinger (1D)} & \pc{0.36\textwidth}{$\displaystyle \frac{\partial^2 \psi}{\partial x^2} = \frac{2 m}{\hbar^{2}} \left(V - E\right) \psi$} &  & \partialmark & \passmark \\
\pc{0.32\textwidth}{Free particle wave} & \pc{0.36\textwidth}{$\displaystyle \frac{\partial^2 \psi}{\partial x^2} = - k^{2} \psi$} &  & \passmark & \passmark \\
\pc{0.32\textwidth}{Harmonic oscillator QM} & \pc{0.36\textwidth}{$\displaystyle \frac{\partial^2 \psi}{\partial x^2} = \left(k^{2} x^{2} - E\right) \psi$} &  & \passmark & \passmark \\
\pc{0.32\textwidth}{Duffing oscillator} & \pc{0.36\textwidth}{$\displaystyle \frac{d^2 x}{d t^2} = - \omega^{2} x - \alpha x^{3}$} &  & \passmark & \passmark \\
\pc{0.32\textwidth}{Pendulum (full nonlinear)} & \pc{0.36\textwidth}{$\displaystyle \frac{d^2 \theta}{d t^2} = - \frac{g}{L} \sin{\left(\theta \right)}$} &  & \partialmark & \partialmark \\
\pc{0.32\textwidth}{Weakly nonlinear oscillator} & \pc{0.36\textwidth}{$\displaystyle \frac{d^2 x}{d t^2} = - \omega^{2} x + \epsilon x^{2}$} &  & \passmark & \passmark \\
\pc{0.32\textwidth}{Van der Pol oscillator} & \pc{0.36\textwidth}{$\displaystyle \frac{d^2 x}{d t^2} = - x + \mu \left(1 - x^{2}\right) \frac{dx}{dt}$} &  & \failmark & \passmark \\
\pc{0.32\textwidth}{Mathieu / parametric} & \pc{0.36\textwidth}{$\displaystyle \frac{d^2 x}{d t^2} = - \omega^{2} \left(1 + \epsilon \cos{\left(\Omega t \right)}\right) x$} &  & \failmark & \passmark \\
\pc{0.32\textwidth}{Relativistic oscillator} & \pc{0.36\textwidth}{$\displaystyle \frac{d^2 x}{d t^2} = -\omega^{2} x \left(1 - \frac{1}{c^{2}}\left(\frac{dx}{dt}\right)^{2}\right)^{\frac{3}{2}}$} &  & \failmark & \partialmark \\
\pc{0.32\textwidth}{Spherical acoustic wave} & \pc{0.36\textwidth}{$\displaystyle \frac{\partial^2 p}{\partial r^2} = - \frac{2}{r}\frac{\partial p}{\partial r} - k^{2} p$} &  & \failmark & \passmark \\
\pc{0.32\textwidth}{String mode (exponential branch)} & \pc{0.36\textwidth}{$\displaystyle \frac{\partial^2 y}{\partial x^2} = \frac{\mu y}{T}$} &  & \passmark & \passmark \\
\pc{0.32\textwidth}{Standing wave on string (spatial mode)} & \pc{0.36\textwidth}{$\displaystyle \frac{\partial^2 y}{\partial x^2} = - \frac{\mu \omega^{2} y}{T}$} &  & \passmark & \passmark \\
\pc{0.32\textwidth}{Capillary rise (planar meniscus)} & \pc{0.36\textwidth}{$\displaystyle \frac{d^2 y}{d x^2} = \frac{\rho g}{\gamma}\, y \left(1 + \left(\frac{dy}{dx}\right)^{2}\right)^{\frac{3}{2}}$} &  & \failmark & \passmark \\
\pc{0.32\textwidth}{Abel equation type I} & \pc{0.36\textwidth}{$\displaystyle \frac{d y}{d x} = x + y^{3}$} &  & \passmark & \passmark \\
\pc{0.32\textwidth}{Riccati equation} & \pc{0.36\textwidth}{$\displaystyle \frac{d y}{d x} = x + y^{2}$} &  & \passmark & \passmark \\
\pc{0.32\textwidth}{Bernoulli equation} & \pc{0.36\textwidth}{$\displaystyle \frac{d y}{d x} = y^{2} + y$} &  & \passmark & \passmark \\
\pc{0.32\textwidth}{Population with crowding} & \pc{0.36\textwidth}{$\displaystyle \frac{d P}{d t} = r P - a P^{2}$} &  & \passmark & \passmark \\
\pc{0.32\textwidth}{Separable product form} & \pc{0.36\textwidth}{$\displaystyle \frac{d y}{d t} = k t y$} &  & \passmark & \passmark \\
\pc{0.32\textwidth}{Separable rational} & \pc{0.36\textwidth}{$\displaystyle \frac{d y}{d x} = \frac{x y}{x^{2} + 1}$} &  & \failmark & \failmark \\
\pc{0.32\textwidth}{Exact, separable} & \pc{0.36\textwidth}{$\displaystyle \frac{d y}{d x} = -\frac{y}{x}$} &  & \partialmark & \passmark \\
\pc{0.32\textwidth}{Prey equation (y=const)} & \pc{0.36\textwidth}{$\displaystyle \frac{d x}{d t} = a x - b x y$} &  & \passmark & \passmark \\
\pc{0.32\textwidth}{Driven damped oscillator (const force)} & \pc{0.36\textwidth}{$\displaystyle \frac{d^2 x}{d t^2} = - a x - b \frac{dx}{dt} + c$} &  & \partialmark & \failmark \\
\pc{0.32\textwidth}{Driven harmonic oscillator} & \pc{0.36\textwidth}{$\displaystyle \frac{d^2 x}{d t^2} = - \omega^{2} x + F_{0} \cos{\left(\Omega t \right)}$} &  & \failmark & \passmark \\
\pc{0.32\textwidth}{Free-particle Schr\"{o}dinger} & \pc{0.36\textwidth}{$\displaystyle i\hbar\,\partial_t\psi = -\frac{\hbar^2}{2m}\,\partial_x^2\psi$} & $\checkmark$ & \passmark & \nodata \\
\pc{0.32\textwidth}{Harmonic oscillator Schr\"{o}dinger} & \pc{0.36\textwidth}{$\displaystyle i\hbar\,\partial_t\psi = -\frac{\hbar^2}{2m}\,\partial_x^2\psi + \tfrac{1}{2}m\omega^2 x^2\psi$} & $\checkmark$ & \passmark & \nodata \\
\pc{0.32\textwidth}{1D Schr\"{o}dinger + potential} & \pc{0.36\textwidth}{$\displaystyle i\hbar\,\partial_t\psi = -\frac{\hbar^2}{2m}\,\partial_x^2\psi + V\psi$} & $\checkmark$ & \passmark & \nodata \\
\pc{0.32\textwidth}{Gross-Pitaevskii (Bose-Einstein condensate)} & \pc{0.36\textwidth}{$\displaystyle i\,\partial_t\psi = -\partial_x^2\psi + g|\psi|^2\psi$} & $\checkmark$ & \passmark & \nodata \\
\pc{0.32\textwidth}{Two-level quantum system} & \pc{0.36\textwidth}{\begin{tabular}[t]{@{}l@{}}$\displaystyle i\dot{c}_1 = E_1 c_1 + V_{12}c_2$\\$\displaystyle i\dot{c}_2 = V_{21}c_1 + E_2 c_2$\end{tabular}} & $\checkmark$ & \passmark & \nodata \\
\pc{0.32\textwidth}{Pauli spin-$\tfrac{1}{2}$ (spin in magnetic field)} & \pc{0.36\textwidth}{\begin{tabular}[t]{@{}l@{}}$\displaystyle i\dot{\psi}_\uparrow = \tfrac{p^2}{2m}\psi_\uparrow + B_z\psi_\uparrow + B_x\psi_\downarrow$\\$\displaystyle i\dot{\psi}_\downarrow = \tfrac{p^2}{2m}\psi_\downarrow - B_z\psi_\downarrow + B_x\psi_\uparrow$\end{tabular}} & $\checkmark$ & \passmark & \nodata \\
\pc{0.32\textwidth}{Time-independent Schr\"{o}dinger} & \pc{0.36\textwidth}{$\displaystyle -\frac{\hbar^2}{2m}\psi'' + V\psi = E\psi$} & $\checkmark$ & \partialmark & \nodata \\
\pc{0.32\textwidth}{Hydrogen radial} & \pc{0.36\textwidth}{$\displaystyle -\frac{\hbar^2}{2m}\!\left(R'' + \frac{2}{r}R'\right) + V_{\text{eff}}R = ER$} & $\checkmark$ & \partialmark & \nodata \\
\pc{0.32\textwidth}{Dirac equation (1+1D)} & \pc{0.36\textwidth}{$\displaystyle i\gamma^\mu\partial_\mu\psi = m\psi$} & $\checkmark$ & \passmark & \nodata \\
\pc{0.32\textwidth}{Nonlinear Schr\"{o}dinger (pulse propagation in fiber)} & \pc{0.36\textwidth}{$\displaystyle i\,\partial_z u + \partial_t^2 u + |u|^2 u = 0$} & $\checkmark$ & \passmark & \nodata \\
\pc{0.32\textwidth}{Complex Ginzburg-Landau (pattern formation)} & \pc{0.36\textwidth}{$\displaystyle \partial_t u = (1{+}ia)\,\partial_x^2 u + u - (1{+}ib)|u|^2 u$} & $\checkmark$ & \passmark & \nodata \\
\pc{0.32\textwidth}{Manakov system (coupled polarisation in fiber)} & \pc{0.36\textwidth}{\begin{tabular}[t]{@{}l@{}}$\displaystyle i\,\partial_z u + \partial_t^2 u + (|u|^2{+}|v|^2)u = 0$\\$\displaystyle i\,\partial_z v + \partial_t^2 v + (|u|^2{+}|v|^2)v = 0$\end{tabular}} & $\checkmark$ & \passmark & \nodata \\
\pc{0.32\textwidth}{Soliton envelope} & \pc{0.36\textwidth}{$\displaystyle i\,\partial_t A + \partial_x^2 A + 2|A|^2 A = 0$} & $\checkmark$ & \passmark & \nodata \\
\pc{0.32\textwidth}{Second harmonic generation (frequency doubling in crystals)} & \pc{0.36\textwidth}{\begin{tabular}[t]{@{}l@{}}$\displaystyle i\,\partial_z A_1 = \kappa\bar{A}_1 A_2$\\$\displaystyle i\,\partial_z A_2 = \kappa A_1^2$\end{tabular}} & $\checkmark$ & \passmark & \nodata \\
\pc{0.32\textwidth}{Parametric amplification (nonlinear optical gain)} & \pc{0.36\textwidth}{$\displaystyle i\,\partial_z A = \kappa\bar{A}\,e^{i\Delta z}$} & $\checkmark$ & \passmark & \nodata \\
\pc{0.32\textwidth}{Complex damped oscillator} & \pc{0.36\textwidth}{$\displaystyle \ddot{z} + 2\gamma\dot{z} + \omega_0^2 z = 0$} & $\checkmark$ & \passmark & \nodata \\
\pc{0.32\textwidth}{Driven complex oscillator} & \pc{0.36\textwidth}{$\displaystyle \ddot{z} + 2\gamma\dot{z} + \omega_0^2 z = Fe^{i\Omega t}$} & $\checkmark$ & \passmark & \nodata \\
\pc{0.32\textwidth}{Coupled complex modes} & \pc{0.36\textwidth}{\begin{tabular}[t]{@{}l@{}}$\displaystyle \dot{z}_1 = i\omega_1 z_1 + \kappa z_2$\\$\displaystyle \dot{z}_2 = i\omega_2 z_2 + \kappa z_1$\end{tabular}} & $\checkmark$ & \passmark & \nodata \\
\pc{0.32\textwidth}{Van der Pol amplitude equation (Stuart--Landau form)} & \pc{0.36\textwidth}{$\displaystyle \dot{A} = (\mu{-}|A|^2)A + i\omega A$} & $\checkmark$ & \passmark & \nodata \\
\pc{0.32\textwidth}{Series circuit phasor (algebraic)} & \pc{0.36\textwidth}{$\displaystyle i\omega L I + RI + \frac{I}{i\omega C} = V$} & $\checkmark$ & \passmark & \nodata \\
\pc{0.32\textwidth}{Gap relaxation (time-dependent Ginzburg--Landau amplitude)} & \pc{0.36\textwidth}{$\displaystyle \dot{\Delta} = g(N_0{-}|\Delta|^2)\Delta$} & $\checkmark$ & \passmark & \nodata \\
\pc{0.32\textwidth}{Landau order parameter (phase transitions)} & \pc{0.36\textwidth}{$\displaystyle \partial_t\Phi = -a\Phi - b|\Phi|^2\Phi + \kappa\,\partial_x^2\Phi$} & $\checkmark$ & \passmark & \nodata \\
\pc{0.32\textwidth}{Josephson junction (superconducting weak link)} & \pc{0.36\textwidth}{\begin{tabular}[t]{@{}l@{}}$\displaystyle \dot{\phi}=V$\\$\displaystyle \dot{V}=-\!\sin\phi - \alpha V + I_{\text{ext}}$\end{tabular}} & $\checkmark$ & \passmark & \nodata \\
\pc{0.32\textwidth}{Stuart-Landau oscillator (limit cycle near bifurcation)} & \pc{0.36\textwidth}{$\displaystyle \dot{A} = (\mu{+}i\omega)A - (1{+}i\beta)|A|^2 A$} & $\checkmark$ & \passmark & \nodata \\
\pc{0.32\textwidth}{Complex Klein-Gordon (relativistic scalar field)} & \pc{0.36\textwidth}{$\displaystyle \partial_t^2\phi - c^2\,\partial_x^2\phi + m^2\phi = 0$} & $\checkmark$ & \partialmark & \nodata \\
Coupled Bose-Einstein condensates (interacting quantum gases) & \begin{tabular}[t]{@{}l@{}}$\displaystyle i\,\partial_t\psi_1 = -\partial_x^2\psi_1 + g_1|\psi_1|^2\psi_1 + \kappa\psi_2$\\$\displaystyle i\,\partial_t\psi_2 = -\partial_x^2\psi_2 + g_2|\psi_2|^2\psi_2 + \kappa\psi_1$\end{tabular} & $\checkmark$ & \passmark & \nodata
\enddata
\end{deluxetable*}

\bibliographystyle{aasjournalv7}
\bibliography{nestynet_paper4}

\end{document}